\documentclass[aps,twocolumn,prd,reprint,superscriptaddress,nofootinbib,floatfix]{revtex4-2}
\RequirePackage[l2tabu, orthodox]{nag}
\usepackage{lipsum}

\usepackage[utf8x]{inputenc}
\usepackage{microtype}

\usepackage{amsmath}
\usepackage{amstext, amssymb, amsthm, amsfonts, slashed}
\usepackage{mathtools}
\usepackage[version=4]{mhchem}
\usepackage{graphicx}
\usepackage{booktabs}
\usepackage{dcolumn}
\usepackage{bm}
\usepackage{dsfont}
\usepackage[usenames,dvipsnames]{xcolor}
\usepackage[normalem]{ulem}

\usepackage{url}
\usepackage[colorlinks=true,urlcolor=blue,linkcolor=blue,citecolor=blue,hypertexnames]{hyperref}
\usepackage[nameinlink]{cleveref}
\usepackage{braket}
\usepackage{simplewick}
\usepackage{float}
\usepackage{subcaption}
\usepackage{physics}

\hypersetup{
	colorlinks,
	linkcolor={blue!75!black},
	citecolor={blue!75!black},
	urlcolor={blue!75!black}
}

\newcommand{\imag}{\text{i}}

\newcommand{\orcid}[1]{\href{https://orcid.org/#1}{\includegraphics[height=1.7ex,width=1.7ex]{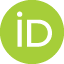}}}

\begin{document}

\title{Self-consistent graviton spectral function in Lorentzian quantum gravity}

\author{Jan~M.~Pawlowski~\orcid{0000-0003-0003-7180}}
\affiliation{Institut für Theoretische Physik, Universität Heidelberg,	Philosophenweg 16, 69120 Heidelberg, Germany}
\affiliation{ExtreMe Matter Institute EMMI, GSI Helmholtzzentrum f\"ur Schwerionenforschung mbH, Planckstr.\ 1, 64291 Darmstadt, Germany}
\author{Manuel~Reichert~\orcid{0000-0003-0736-5726}}
\affiliation{Department  of  Physics  and  Astronomy,  University  of  Sussex,	Brighton,  BN1	9QH,  U.K.}
\author{Jonas~Wessely~\orcid{0009-0006-8700-104X}}
\affiliation{Institut für Theoretische Physik, Universität Heidelberg, Philosophenweg 16, 69120 Heidelberg, Germany}

\begin{abstract}
	We present the first fully self-consistent computation of the graviton spectral function in quantum gravity, using the spectral renormalisation group for gravity put forward in \cite{Fehre:2021eob} within a physical mass-shell renormalisation scheme. Here, self-consistency refers to the fact that the full non-perturbative spectral function is used in the diagrams, including the scattering continuum. We find a positive graviton spectral function with a massless one-graviton peak and a multi-graviton continuum with a close-to-quadratic spectral decay in the ultraviolet. Within the physical on-shell renormalisation scheme, the graviton satisfies the sum rule of an asymptotic state and features a unit total spectral weight. We briefly discuss the implications of the physical formulation for the computation of scattering processes and investigations of unitarity in asymptotically safe quantum gravity.
\end{abstract}

\maketitle

\section{Introduction}
Over the past three decades, asymptotically safe gravity \cite{Weinberg:1980gg, Reuter:1996cp} has grown into a serious candidate for an ultraviolet (UV) complete theory of quantum gravity. Most of the progress has been achieved within the Euclidean version of the functional renormalisation group (fRG), for recent reviews see \cite{Bonanno:2020bil, Dupuis:2020fhh, Pawlowski:2020qer, Knorr:2022dsx, Eichhorn:2022gku, Morris:2022btf, Wetterich:2022ncl, Martini:2022sll, Saueressig:2023irs, Pawlowski:2023gym, Platania:2023srt, Bonanno:2024xne, Reichert:2020mja}. Test of unitarity and causality are amongst the most pressing open questions, for first steps within the fRG towards Lorentzian quantum gravity see \cite{Fehre:2021eob, Bonanno:2021squ, Kher:2025rve, Baldazzi:2018mtl, Baldazzi:2019kim, DAngelo:2022vsh, Banerjee:2022xvi, DAngelo:2023tis, DAngelo:2023wje, Banerjee:2024tap, Thiemann:2024vjx, Ferrero:2024rvi, DAngelo:2025yoy} as well as fRG approaches based on ADM-type decompositions \cite{Knorr:2018fdu, Eichhorn:2019ybe, Knorr:2022mvn,  Saueressig:2023tfy, Korver:2024sam, Saueressig:2025ypi} and the discussion of various aspects of unitarity \cite{Platania:2020knd, Platania:2022gtt, Draper:2020bop, Knorr:2020bjm, Knorr:2022lzn, Pastor-Gutierrez:2024sbt, Knorr:2024yiu, Eichhorn:2024wba}.

In \cite{Fehre:2021eob}, the first Lorentzian RG approach to quantum gravity was set up, based on the non-perturbative functional spectral approach put forward in \cite{Horak:2020eng, Fehre:2021eob, Braun:2022mgx} and applied to various quantum field theories, including gauge theories in \cite{Horak:2021pfr, Horak:2022myj, Horak:2022aza, Horak:2023hkp, Pawlowski:2024kxc}. As a first successful application, the graviton spectral function and the graviton propagator in the complex frequency plane were computed in \cite{Fehre:2021eob}, in good agreement with the reconstruction results obtained in \cite{Bonanno:2021squ}. More recently also matter spectral function in quantum gravity have been computed \cite{Kher:2025rve}. These exciting results provide direct access to timelike scattering processes and, hence, to the question of unitarity in asymptotically safe quantum gravity, see \cite{Pastor-Gutierrez:2024sbt} for a first computation.

In the present work, we concentrate on two novel properties that accommodate both technical improvements as well as conceptual progress: \\[-2ex]

Firstly, we use a self-consistent approximation that includes the full feedback of the spectral function into the flow. Within this approximation, the full leading order for sub-Planckian momenta is incorporated and leads to the known exact infrared (IR) limit of the graviton spectral function in this regime. Moreover, we also show that the sub-leading nature of the scattering continuum in the flow is well-supported, which facilitates further computations.

Secondly, we use the on-shell renormalisation scheme set up in \cite{Horak:2020eng, Braun:2022mgx, Horak:2023hkp}. In this physical renormalisation scheme, the renormalisation constants are set to unity on-shell, which leads to a classical on-shell dispersion. This amounts to a significant technical as well as numerical simplification, and is also linked to an optimal physical expansion scheme in the fRG. Importantly, this flowing re-normalisation of the graviton fluctuation field leads to a \textit{positive} and \textit{normalisable} graviton spectral function which satisfies the spectral sum rule of an asymptotic state. This entails a unity spectral weight that is related to probability conservation and unitarity of metric gravity, as well as implying canonical commutation relations of the graviton field operator. In summary, we put forward a reformulation of quantum gravity in terms of close-to-physical graviton fields.

\section{Renormalised spectral flows for the graviton propagator}
\label{sec:GravitonspectralFlow}
The computation of the graviton spectral function builds upon the spectral RG for Lorentzian gravity put forward in \cite{Fehre:2021eob}, combined with the on-shell renormalisation scheme in spectral functional approaches, \cite{Horak:2020eng, Braun:2022mgx, Horak:2023hkp}. Specifically, we use the flow equations for the graviton propagator derived in \cite{Fehre:2021eob} and solve them self-consistently within on-shell renormalisation, feeding back the full graviton spectral function. The implementation of the spectral RG in \cite{Fehre:2021eob} for the computation of the graviton spectral function relies on the -non-trivial- existence of the Källén-Lehmann (KL) spectral representation for the graviton propagator, which also underlies the present application. In the following, we briefly review the relevant equations and derivations; for more details, we refer to \cite{Fehre:2021eob, Braun:2022mgx}.

\subsection{Renormalised spectral flows}
\label{sec:SpectralFlows}
We study metric quantum gravity where the dynamical metric field $g_{\mu\nu}$ carries the fundamental gravitational degrees of freedom. We use a linear split of the full metric into a flat Minkowski background $\eta_{\mu\nu}$ and a fluctuation field $h_{\mu\nu}$,
\begin{align}
	g_{\mu\nu}=\eta_{\mu\nu} + \sqrt{32 \pi \,G_\text{N}}\,\left( \sqrt{Z_h}  h_{\mu\nu} \right)\,,
	\label{eq:metric_split}
\end{align}
with the Newton constant $G_\text{N}$ and the wavefunction renormalisation $Z_h$ of the graviton field $h_{\mu\nu}$. The Newton coupling ensures that the fluctuation field $h_{\mu\nu}$ has mass dimension one, typical for a bosonic field, and the combination $\sqrt{Z_h}\, h$ is a convenient choice since it is RG invariant as discussed later. Studying the correlation functions of the graviton field $h_{\mu\nu}$ is underlying the fluctuation approach to quantum gravity, see \cite{Pawlowski:2020qer, Pawlowski:2023gym} for more details.

While the split \labelcref{eq:metric_split} singles out the Minkowski metric as background, the theory is independent of the choice of the background metric. Still, the underlying expansion about the flat background may suffer from slow convergence compared to an on-shell background, for more details on background independence in the fluctuation approach, see again \cite{Pawlowski:2020qer, Pawlowski:2023gym}. The Minkowski metric is a particularly convenient choice as it allows for a straightforward definition of the spectral function as well as the discussion of spectral sum rules. These properties are more intricate within non-trivial backgrounds that may, in particular, also serve as dynamical sources or sinks, hence influencing the spectral sum rules as well as the existence of the KL representation itself.

Now we briefly discuss the spectral flow setup. We use the spectral fRG flow for the scale-dependent effective action $\Gamma[\Phi]$ of quantum gravity. The set of fields $\Phi$ includes the graviton $h$, and the gauge-fixing ghost and antighost $c,\,\bar c$,
\begin{align}
	\Phi = \left(h_{\mu\nu}\,,\, c_\mu\,,\, \bar{c}_\mu\right),
	\label{eq:SuperField}
\end{align}
where we suppressed the dependence on the background metric. The effective action satisfies the renormalised Callan-Symanzik (CS) flow equation \cite{Fehre:2021eob, Braun:2022mgx}, based on~\cite{Symanzik:1970rt},
\begin{align}
	\partial_t \Gamma[\Phi] = \frac{1}{2} \Tr [ \mathcal{G}[\Phi]\, \partial_t R_{\textrm{\tiny{CS}}}] - \partial_t S_{\textrm{ct}} \,,
	\label{eq:RenCS-flow}
\end{align}
where the trace denotes a trace over all internal, spacetime, and momentum indices of the superfield, including a relative minus sign for fermionic fields. Note that $\Gamma$ is the full effective action including the cutoff term. This is different from the common split used in the Wetterich equation \cite{Wetterich:1992yh}, where the cutoff term is subtracted in the definition of $\Gamma$. \Cref{eq:RenCS-flow} depends on the full field-dependent propagator ${\mathcal G}[\Phi]$ of the fields with
\begin{align}
	{\mathcal G}[\Phi] = \frac{1}{\Gamma^{(2)} } \,,
	\label{eq:def-G}
\end{align}
where $\Gamma^{(2)}_{\Phi_i\Phi_j}$ or $\Gamma^{(\Phi_i\Phi_j)}$ stand for the second derivative of the effective action with respect to the fields $\Phi_i$ and $\Phi_j$. Note that in \labelcref{eq:RenCS-flow,eq:def-G} and in the following all quantities are $k$ dependent, which is implicitly understood.

The CS regulator $R_{\textrm{\tiny{CS}}}$ is a matrix in field space, and is diagonal for bosonic fields and symplectic for fermionic ones. Its entries are given by
\begin{align}
	R_{hh}       & = Z_h k^2\,,
	             &
	R_{\bar{c}c} & = Z_c k^2\,,
	\label{eq:Regulators}
\end{align}
for the fluctuation graviton and the ghost, respectively. The prefactors $Z_h$ and $Z_c$ make these regulators RG adapted, see \cite{Pawlowski:2005xe}. This entails that the underlying RG or reparametrisation invariance of the theory is kept intact for $k\neq 0$.

The momentum independence of this regulator preserves the analytic properties in the complex plane, which is related to unitarity and causality. It comes at the price of the flowing counter term $\partial_t S_\textrm{ct} $ in \labelcref{eq:RenCS-flow}, which ensures manifest finiteness of the flow. It is worth emphasising that this term is not added ad hoc, but originates within a controlled limit of manifestly finite functional flows with momentum-dependent regulators, see \cite{Braun:2022mgx}. In \labelcref{eq:RenCS-flow}, the flowing counter term also includes the trivial running of the cutoff term.

These flows are variants of the standard Wetterich equation \cite{Wetterich:1992yh}, see also~\cite{Ellwanger:1993mw, Morris:1993qb}. Its practical feasibility has been proven in applications to quantum gravity  \cite{Fehre:2021eob}, as well as scalar theories~\cite{Horak:2023hkp, Kockler:2025kdt}. Note also that the absorption of the cutoff term into the effective action is the 'natural' choice if using on-shell renormalisation, see \cite{Horak:2023hkp,  Kockler:2025kdt}: the CS-cutoff term is simply a contribution to the mass term and the CS flow is one in the space of theory. In contradistinction to momentum-dependent regulators, the respective theories are physical but feature different masses. In gravity, the situation is more subtle as different graviton mass parameters relate to different cosmological constants; for more details, see \cite{APRW}.

\begin{figure}
	\includegraphics[width=\linewidth]{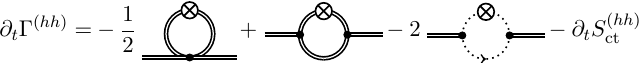}
	\caption{Flow equation for the graviton two-point function. The lines stand for full propagators of gravitons (straight double lines) and ghosts (dotted lines), and the vertices are the full vertices derived from the effective action. The cutoff insertion $\partial_t R_{\textrm{\tiny{CS}}}$ is depicted as a circled cross. \hspace*{\fill}}
	\label{fig:diagrams}
\end{figure}

We proceed with the discussion of the approximation of the effective action used for our computations. We use an RG-invariant vertex expansion scheme \cite{Christiansen:2014raa, Christiansen:2015rva, Denz:2016qks, Pawlowski:2020qer, Pawlowski:2023gym, Ihssen:2024miv}, and the effective action reads schematically
\begin{subequations}
	\label{eq:RG-InvScheme}
	\begin{align}
		\Gamma[\Phi] =  \sum_{\boldsymbol{n}} \frac{1}{\boldsymbol{n}!} \int\bar \Gamma^{(\Phi_{i_1}\cdots \Phi_{i_n})}\,\bar\Phi_{i_1}\cdots \bar \Phi_{i_n}\,,
		\label{eq:GammaRGINv}
	\end{align}
	where $\boldsymbol{n}=(n_h,n_c,n_{\bar c})$ and $\boldsymbol{n}!$ takes into account the multiplicities of the different species of fields. The integral stands for $n_h+n_c+n_{\bar c}$ momentum integrations, and we have suppressed the momentum arguments of both the vertices and the fields. The RG-invariant vertices and fields are defined by
	\begin{align}
		\bar \Gamma^{(\Phi_{i_1}\cdots \Phi_{i_n})} & = \frac{\Gamma^{(\Phi_{i_1}\cdots \Phi_{i_n})} }{Z_{\Phi_{i_1}}^{1/2}\cdots Z_{\Phi_{i_n}}^{1/2}} \,,
		                                            &
		\bar \Phi_{i}                               & = Z_{\Phi_{i}}^{1/2}  \Phi_{i}\,.
		\label{eq:GammaBarn}
	\end{align}
\end{subequations}
This factorisation improves the convergence of expansion schemes with a derivative expansion or a partial momentum dependence.

In the present application, we use an approximation of the effective action, which is a combination of a full, cutoff- and momentum-dependent graviton two-point function and an Einstein-Hilbert type part for the vertices. Written in terms of the RG-invariant vertex expansion \labelcref{eq:RG-InvScheme}, the approximation is given by
\begin{align}\nonumber
	\Gamma[\phi] = & \,\frac12 \int_p \bar h^{\mu\nu}\,\bar \Gamma^{(hh)}_{\mu\nu\rho\sigma}(p) \,\bar h^{\rho\sigma}  + S_\text{gf} + S_\text{gh}                                 \\[1ex]
	               & \,+  \sum_{|\boldsymbol{n}|=3} \frac{1}{\boldsymbol{n}!} \int S_\textrm{EH}^{(\bar\Phi_{i_1}\cdots \bar \Phi_{i_n})}\,\bar\Phi_{i_1}\cdots \bar \Phi_{i_n}\,,
	\label{eq:GammaPhi}
\end{align}
with the flowing Einstein-Hilbert action
\begin{align}
	S_{\text{EH}}=   \frac{1}{16\pi G_{\text{N},k}} \int \! \mathrm  d^4x\sqrt{g}\left(R-2\Lambda_k\right)  \,,
	\label{eq:FlowingSEH}
\end{align}
with $\sqrt{g} = \sqrt{|\operatorname{det}g_{\mu\nu}|}$, the running Newton coupling $G_{\text{N},k}$ and the running cosmological constant $\Lambda_k$. The gauge fixing used in \labelcref{eq:GammaPhi} is of the de-Donder type, see \Cref{app:GaugeFixing}, and we use the harmonic gauge with
\begin{align}
	\alpha=\beta = 1\,,
	\label{eq:HarmonicGauge}
\end{align}
which is singled out by spectral considerations, see \cite{Fehre:2021eob, ALR}. Moreover, we only consider a vanishing physical cosmological constant,
\begin{align}
	\Lambda_{k=0}=0\,.
	\label{eq:Lambdaphys0}
\end{align}
We shall extend \labelcref{eq:Lambdaphys0} to $k\neq 0$ for computational convenience, that is $\Lambda_{k}=0$.

\subsection{Spectral flows for the graviton spectral function}
\label{sec:SpectralFlowsGraviton}

We now discuss the numerical computation of the Lorentzian graviton two-point function $\Gamma^{(2)}_{hh}(p)$ from its flow depicted in \Cref{fig:diagrams}. The diagrams depend on the full graviton propagator, whose momentum- and frequency dependence is computed using its spectral representation. The graviton propagator is composed of five tensor structures, and we are studying the scalar propagator part $G_{hh}$ of the traceless-transverse mode,
\begin{align}
	{\mathcal{G}}^{\mu\nu\rho\sigma}_{hh}(p) =  G_{hh}(p)
	\,	\Pi_\text{TT}^{\mu\nu\rho\sigma}(p) + \dots\,,
	\label{eq:GTT}
\end{align}
where $\Pi_\text{TT}$ is the transverse-traceless projection operator, defined in \labelcref{eq:Pi_TT}, and the dots indicate the other modes of the graviton.  We identify their scalar parts with $G_{hh}$, and \labelcref{eq:GTT} also defines implicitly the two-point function $\Gamma^{(2)}_\text{TT}  =G_{hh}^{-1}$. For a discussion of the full graviton propagator and the gauge dependence of the spectral representation, see~\cite{ALR}.

For the scalar propagator function $G_{hh}(p)$, we use the KL spectral representation,
\begin{align}
	G_{hh}(p^2) = \int_\lambda \frac{ \rho_h(\lambda)}{\lambda^2 + p^2} \,, \qquad \int_\lambda = \int_{0}^\infty \frac{\mathrm{d}\lambda\, \lambda}{\pi}\,,
	\label{eq:KL}
\end{align}
which gives us access to the graviton propagator in the complex frequency plane, subject to a given graviton spectral function $\rho_h$.

The spectral function can be extracted from the imaginary part of the retarded graviton propagator,
\begin{align}
	\rho_h(\omega) = 2\, \text{Im} \, G_{hh}\left(p^2 \to -(\omega_+)^2\right) \,,
\end{align}
where the subscript $_+$ denotes the limit $f(\omega_+) = \lim_{\epsilon \searrow 0} f(\omega + \imag \epsilon)$.

The spectral function is parametrised by
\begin{align}
	\rho_h(\lambda) = \frac{1}{Z_h} \Big[ 2\pi \delta(\lambda^2 - m_h^2) + \theta(\lambda^2 - 4 m_h^2) f_h(\lambda) \Big]\,.
	\label{eq:rhoh-para}
\end{align}
The $\delta$-function implements a single particle peak with the flowing graviton pole $m_h^2$ with $m^2_h=0$ at $k=0$. For the flat Minkowski background, this property is obtained with the restriction to a vanishing cosmological constant. For a first discussion of the changes and their interpretation for $\Lambda\neq 0$ see \cite{Fehre:2021eob}. The continuum $f_h$ accommodates higher scattering states, the multi-graviton and graviton–ghost processes.

Finally, for fields that are directly linked to asymptotic states, the spectral function satisfies a spectral sum rule,
\begin{align}
	\int_\lambda\,\rho(\lambda)=1\,.
	\label{eq:Sumrule}
\end{align}
This is related to probability conservation and unitarity in the given theory, if the spectral function is positive semidefinite, and \labelcref{eq:Sumrule} implies canonical commutation relations; for more details, see \cite{Pawlowski:2023gym} and for a discussion in perturbatively controlled gauge theories, see \cite{Kluth:2022wgh}. While the graviton is not directly linked to an asymptotic state, we shall see that in the present setup with on-shell renormalisation we obtain a normalisable positive spectral function $\rho_h(\lambda)\geq  0$. We emphasise that the \textit{flowing} on-shell renormalisation implements a \textit{momentum-dependent} rescaling of the fluctuation graviton in a consistent way, and it is this momentum-dependent rescaling that leads to the normalisable spectral function.

\subsection{On-shell renormalisation}
\label{sec:OnShellRenorm}

The normalisation $Z_h$ is fixed in the on-shell renormalisation scheme, already applied in \cite{Horak:2023hkp, Kockler:2025kdt}. In this scheme, the running pole mass is identified with the RG scale and the on-shell wave function $Z_h$ is set to unity,
\begin{align}
	m_h^2 & =k^2\,,
	      &
	Z_h   & =1\,,
	\label{eq:OnshellConds}
\end{align}
which is implemented via
\begin{align}\nonumber
	\Gamma^{(2)}(p^2 = -k^2)                & = 0 \,, \\[1ex]
	\partial_{p^2} \Gamma^{(2)}(p^2 = -k^2) & = 1\,.
	\label{eq:onshell-conditionG}
\end{align}
This allows us to extract the (integrated) coefficients of the counter-term action from the diagrams evaluated on-shell. Note also that \labelcref{eq:Sumrule} is readily implemented in a final step, subject to the existence of a finite spectral weight.

Let us illustrate the implications of \labelcref{eq:onshell-conditionG} with the traceless-transverse two-point function in its common parametrisation,
\begin{align}
	\Gamma_\text{TT}^{(2)}(p) = Z_h(p)\Bigl[ p^2 + k^2(1 + \mu_h) \Bigr]\,.
	\label{eq:G2TTCommon}
\end{align}
Here, $\mu_h$ is the graviton mass parameter. Note again that $\Gamma^{(2)}$ includes the CS-cutoff term $Z_h k^2$, see \labelcref{eq:RenCS-flow}. \Cref{eq:onshell-conditionG} implies
\begin{align}
	\mu_h           & = 0\,,
	                &
	Z_h(p^2 = -k^2) & = 1\,.
	\label{eq:GravParameters}
\end{align}
\Cref{eq:OnshellConds} adjusts a classical on-shell dispersion and, as part of a ground state expansion, potentially optimises the physics content and convergence of a given expansion scheme, see also \cite{Ihssen:2023nqd, Ihssen:2024ihp}.

The on-shell renormalisation conditions \labelcref{eq:onshell-conditionG} can be recast in terms of flowing renormalisation conditions,
\begin{align}\nonumber
	\partial_t \Gamma_\text{TT}^{(2)} (p^2 = -k^2)                 & = 2 k^2 \,, \\[1ex]
	\partial_{p^2}	\partial_t  \Gamma_\text{TT}^{(2)} (p^2 = -k^2) & = 0 \,,
	\label{eq:onshell-conditiondG}
\end{align}
where the $t$-derivatives are taken at fixed momentum argument. In particular, this implies
\begin{align}
	\eta_h(p^2=-k^2)  = 0 \,.
	\label{eq:onshell-conditioneta}
\end{align}
%
%
Specifically, $Z_h=1$, implemented via \labelcref{eq:onshell-conditioneta}, implies that $\Gamma^{(n)}=\bar\Gamma^{(n)}$ which underlines the physical nature of on-shell renormalisation. This flowing renormalisation accounts for a momentum-dependent rescaling of the fluctuation graviton: the renormalisation condition is applied at all $p^2=-k^2$ with $k\in [0,\infty )$, that is, all momenta. The current computational scheme also uses $\Lambda_k=0$ in the vertices, see \labelcref{eq:GammaPhi}.

\subsection{Computational setup}
\label{sec:CompSetup}

We proceed with the flow of the RG-invariant graviton two-point function $\bar\Gamma^{(2)}$. In on-shell renormalisation, it is given by that of $\Gamma^{(2)}$,
\begin{align}
	\partial_t {\Gamma}^{(hh)}_\text{TT} =  {\textrm{Flow}}^{(hh)}_\text{tadpole} +{\textrm{Flow}}^{(hh)}_\text{3-point} +  {\textrm{Flow}}^{(hh)}_\text{ghost} \,.
	\label{eq:FlowG2}
\end{align}
where the terms on the right-hand side are the diagrammatic part of the flow in \Cref{fig:diagrams}, including the counter terms. For the explicit expressions in terms of KL representations, see \labelcref{eq:GravDiags}.

The approximation is closed with the $\beta$-function of the Newton coupling. It would be desirable to compute it on-shell, extending the on-shell renormalisation at $p^2=-k^2$ to the coupling, which will be discussed in \cite{APRW}. In the present work, we use existing flows at $p=0$ derived with a Litim regulator from the graviton three-point function \cite{Christiansen:2015rva, Denz:2016qks}. In the harmonic gauge \labelcref{eq:HarmonicGauge} and with the on-shell conditions $\mu_h=\eta_h=0$ \labelcref{eq:GravParameters}, we arrive at
\begin{align}
	\partial_t g & = 2g - \frac{2499}{380\pi}g^2\,,
	             &
	g            & =G_{\text{N},k} k^2\,.
	\label{eq:betafunction_g}
\end{align}
The Newton coupling has a UV attractive fixed point at
\begin{align}
	g^* = \frac{760\pi}{2499} \approx 0.955\,.
	\label{eq:gFPvalue}
\end{align}
We expect that using a different cutoff function and different renormalisation conditions leads to mild variations of the beta function of the Newton coupling. The structural analysis of the present work relies only on the asymptotic behaviour of the Newton coupling, and changes of fixed-point value only lead to quantitative changes.

\begin{figure*}
	\centering
	\begin{subfigure}[t]{0.48\linewidth}
		\centering
		\includegraphics[width=\textwidth]{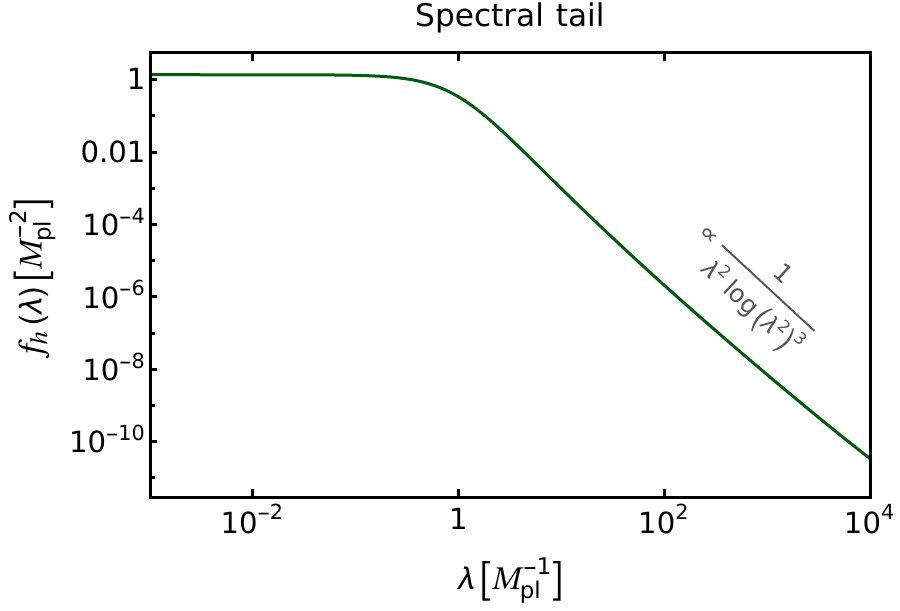}
		\caption{Scattering continuum of the propagator spectral function, with its analytically calculated decay behaviour.  \hspace*{\fill}}
		\label{fig:spectail_graviton}
	\end{subfigure}\hfill
	\centering
	\begin{subfigure}[t]{0.48\linewidth}
		\centering
		\includegraphics[width=\textwidth]{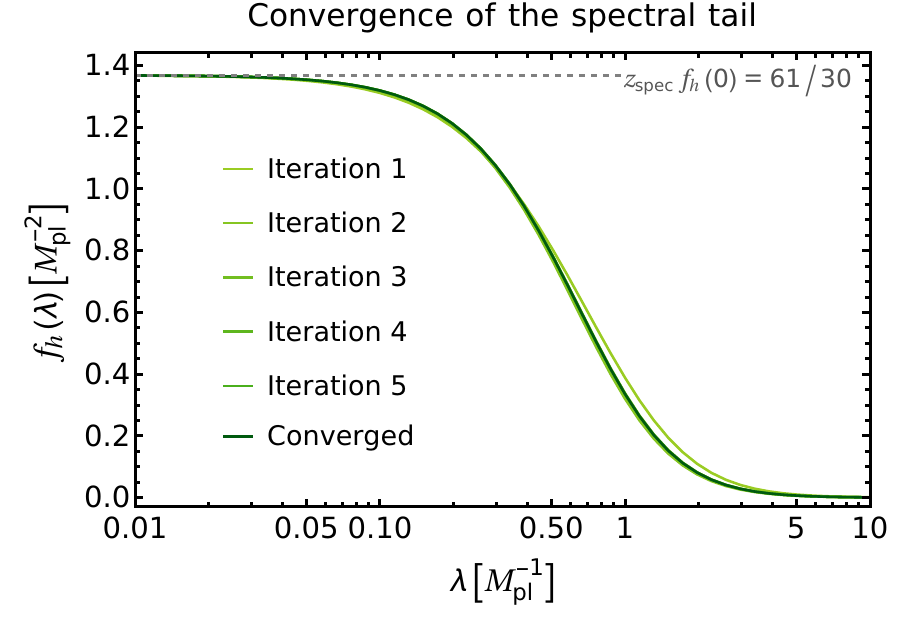}
		\caption{Convergence of the iterative solution, the universal one-loop IR value is indicated in grey. \hspace*{\fill}}
		\label{fig:grav_convergence}
	\end{subfigure}
	\caption{Normalised scattering continuum of the propagator spectral function measured in units of $M_\text{pl}$. \hspace*{\fill} }
	\label{fig:GravResults}
\end{figure*}
The beta function \labelcref{eq:betafunction_g} is readily integrated and yields a simple trajectory for the Newton coupling:
\begin{align}
	g(k) = \frac{g^* k^2}{k^2 + g^* M_\text{pl}^2}\,,
\end{align}
with the fixed-point value $g^*$ in \labelcref{eq:gFPvalue} and the Planck mass $M_\text{pl}$. This concludes our setup and leaves us with the remaining task to integrate the spectral flow of the traceless-transverse part of the graviton two-point function.

For this purpose, we use the integrated flow and solve the respective integral equation for $\rho_{h}(k,\lambda)$. Specifically, we focus on the imaginary part of the diagrams, as the renormalised real part is easily computed with suitably subtracted Kramers-Kronig relations. We solve the integrated CS equation iteratively for the graviton spectral function. After inserting the spectral representation \labelcref{eq:KL} for each line in \Cref{fig:diagrams}, the imaginary part of the RG-invariant graviton two-point function is given by
\begin{align}\nonumber
	\text{Im} \, \Gamma_\text{TT}^{(hh)} (k,\omega) = & \int_k^{\omega/2} \frac{\mathrm dk'}{k'} \int_{\lambda_1,\lambda_2} \bigg\lbrace                                            \\[2ex]
	                                                  & \hspace{-1.5cm}\rho^{(2)}_h(k',\lambda_1)\rho_h(k',\lambda_2)\, \text{Im}\,D(k',\omega,\lambda_1,\lambda_2) \bigg\rbrace\,,
	\label{eq:ImGammaGrav}
\end{align}
where $D(k,\omega,\lambda_1,\lambda_2)$ is the sum of ghost and graviton polarisations, see \labelcref{eq:ImGhostpol,eq:ImGravPol,eq:defImD}. Note that the $k$ integral in \labelcref{eq:ImGammaGrav} for a given frequency $\omega$ only extends to $\omega/2$, since the imaginary part of the flow is proportional to $\theta(\omega - 2k)$.

In \labelcref{eq:ImGammaGrav} we used a generalised spectral representation for the regulator line, which carries the squared propagator,
\begin{subequations}
	\label{eqs:Gsquarerep}
	\begin{align}
		G_{hh}^2(p) = \int_\lambda \frac{\rho_h^{(2)}(\lambda)}{(p^2 + \lambda^2)^2}\,,
		\label{eq:Gsquarerep_int}
	\end{align}
	where the real part of the squared propagator is defined as the Hadamard finite part of the integral. This yields the associated spectral function
	\begin{align}
		\partial_{\omega^2} \rho_h^{(2)}(\omega) = \text{Im}\, G_{hh}^2(p \to -\imag \omega^+)\,.
		\label{eq:defrhoh2}
	\end{align}
\end{subequations}
The squared kernel \labelcref{eq:defrhoh2} leaves us with one spectral integral less, and hence reduces the computational costs, see also \Cref{sec:higherorderkernels} for details.

To extract the spectral function from the left-hand side of \labelcref{eq:ImGammaGrav}, we use the Kramers-Kronig relations, see \labelcref{eq:KramersKronig}, instead of computing the real part directly from the diagrams. This approach is computationally substantially cheaper as it requires only a single (principal value) integral instead of a potentially two-dimensional spectral one. Moreover, it also allows for a convenient implementation of the renormalisation conditions \labelcref{eq:onshell-conditionG}.

Finally, the spectral tail in \labelcref{eq:KL} is given by
\begin{align}
	f_h(k,\omega) = \frac{-2\, \text{Im} \, \Gamma_\text{TT}^{(hh)} (k,\omega)}{ \left[ \text{Im} \, \Gamma_\text{TT}^{(hh)} (k,\omega) \right]^2 + \left[ \text{Re} \, \Gamma_\text{TT}^{(hh)} (k,\omega) \right]^2 }\,,
	\label{eq:f_h_extraction}
\end{align}
for $\omega > 2k$. This closes our system of equations, which we solve via iteration. For numerical details, see \Cref{app:numericaldetails}.

\begin{figure*}
	\includegraphics[width=0.455\textwidth]{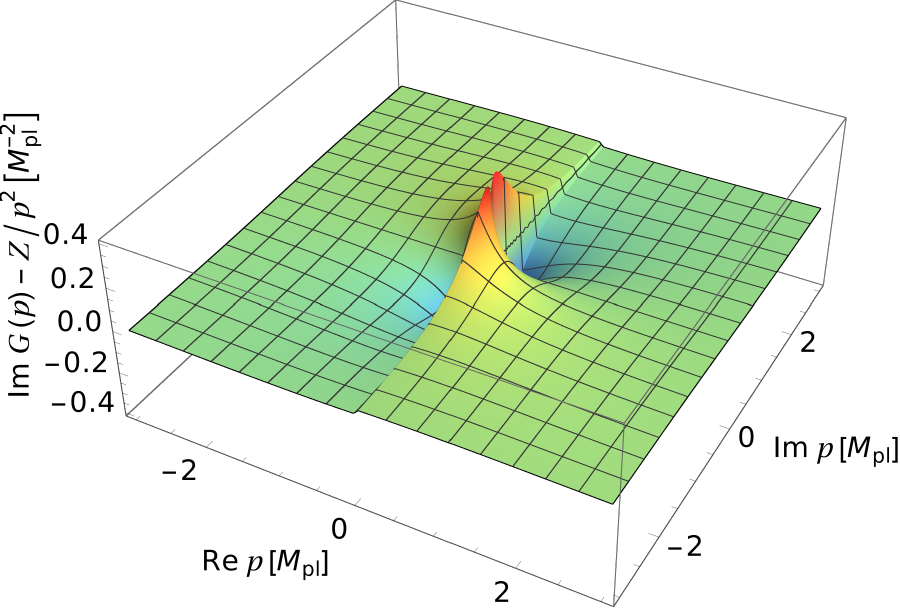}
	\hfill
	\includegraphics[width=0.455\textwidth]{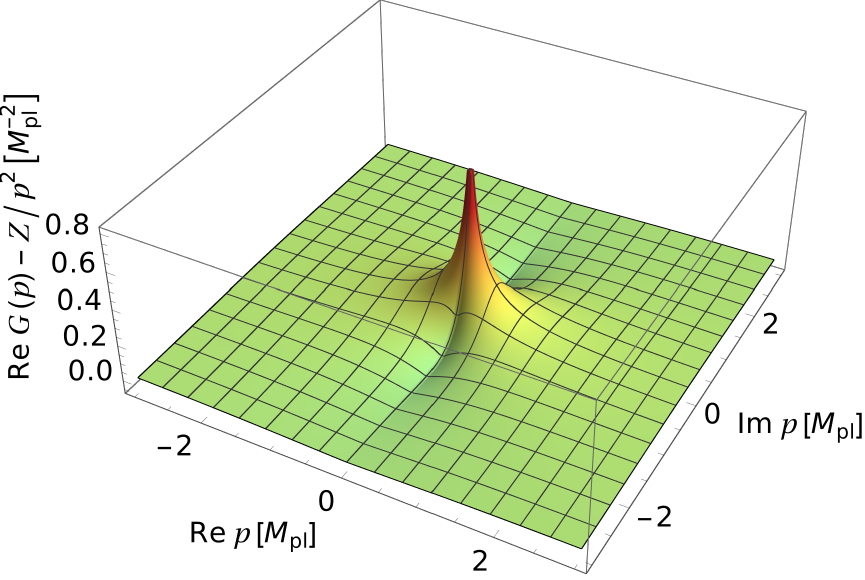}
	\caption{Subleading part of the graviton propagator in the complex plane. It features a cut on the real axis, visible as discontinuity in the imaginary part (left). The cut approaches a constant at the origin, where it flips sign. The corresponding logarithmic divergence is shown in the real part (right). \hspace*{\fill} }
	\label{fig:3dgravprop}
\end{figure*}

\section{Results}
\label{sec:GravResults}

The physical solution for the graviton spectral function is obtained at $k=0$, $\rho(\lambda)=\rho(k=0,\lambda)$. It is obtained from the iterative solution of the integrated CS equation. We show its scattering tail (up to a global rescaling) in \Cref{fig:GravResults}, where we also indicate the analytic IR and UV behaviour. We start with the discussion of the latter, as it has important consequences for the total spectral weight.  For asymptotically large values, the decay of the tail follows a simple behaviour,
\begin{align}
	f_h(\lambda \to \infty) = \frac{c_h^{\textrm{UV}}}{\lambda^2 \log^3(\lambda^2)}\,,
	\label{eq:gravspecdecay}
\end{align}
which is also indicated in \Cref{fig:spectail_graviton}. While the prefactor $c_h^{\textrm{UV}}$ can only be determined numerically, the emergence of the analytic $\lambda$-dependence can be computed analytically if the
scattering tail $f_h$ is not fed back in the flow. Note that this does not affect the decay behaviour itself but only the value of the prefactor. The respective computation has been deferred to \Cref{sec:UV-limit}.

With the decay \labelcref{eq:gravspecdecay}, the spectral integral in the sum rule \labelcref{eq:Sumrule} is finite, and we find
\begin{align}
	z_\textrm{spec} = 1+ \int_\lambda \, \lambda\,f_h(\lambda)\approx 1.486\,.
	\label{eq:ZhNormalised}
\end{align}
Now we define
\begin{align}
	\rho_h^{(\textrm{ph})} = \frac{1}{z_\textrm{spec} }\,\rho_h\,,\qquad \textrm{with} \qquad 	h_{\mu\nu}^{(\textrm{ph})} = h_{\mu\nu}/z^{1/2}_\textrm{spec}\,.
	\label{eq:hrhoPhys}
\end{align}
where $h_{\mu\nu}$ is the fluctuation field with the on-shell renormalisation \labelcref{eq:OnshellConds}. The physical spectral function $\rho_h^{(\textrm{ph})} $ satisfies the sum rule \labelcref{eq:Sumrule},
\begin{align}
	\int_\lambda\,\rho_h^{(\textrm{ph})}(\lambda)=1\,.
	\label{eq:Sumrulerhohphys}
\end{align}
The sum rule \labelcref{eq:Sumrulerhohphys}, or rather a unity total spectral weight, is one of the properties that follow for quantum fields that define an asymptotic state. Put differently, if the fluctuating graviton field operator $\hat h$ would define an asymptotic one-graviton state with $| h \rangle = \hat h|\Omega \rangle $, where $|\Omega \rangle $ is the vacuum state, the satisfaction of the sum rule would follow straightforwardly. A further necessary condition is $\rho_h\geq 0$, which also holds. We conclude that the fluctuating graviton with on-shell renormalisation gets as close as possible to a physical field. We hasten to add that the respective states are not diffeomorphism invariant and hence are not part of the physical Hilbert space. However, the above properties suggest that on-shell renormalisation arranges for 'optimal' building blocks for expansion schemes in asymptotically safe gravity. This also answers the question concerning the overall normalisation of the tail $f_h$ in \Cref{fig:GravResults}, where we show the spectral function $\rho_h^{(\textrm{ph})}$ with the unit spectral weight \labelcref{eq:Sumrulerhohphys}. This facilitates access to the spectral weight of the tail.

We proceed with a discussion of the IR limit of the spectral function, or rather the scattering tail, which is indicated in \Cref{fig:grav_convergence}. In the IR for $\lambda\to 0$, the Newton coupling $G_\text{N}$ settles at its IR value ${{G_{\text{N},k=0}=1/M_\textrm{pl}^2}}$ and all propagators are dominated by their classical values. Hence, quantum corrections to the propagator take a universal form at $k=0$,
\begin{subequations}
	\label{eq:universalIR}
	\begin{align}
		G^{(\textrm{ph})}_{hh} =\frac{1}{z_\text{spec}}\left(\frac{1}{p^{2}} -\,  A_h \log(p^2) + \text{sub-leading}\right)\!,
		\label{eq:universalIR1}
	\end{align}
	with the scheme-independent but gauge-dependent coefficient \cite{Capper:1979ej, Bonanno:2021squ},
	\begin{align}
		A_h = \frac{61}{60\pi} \approx 0.32\,.
	\end{align}
\end{subequations}
This translates into a constant onset of the scattering tail of $\rho_h=z_\text{spec}\,\rho^{(\textrm{ph})}$ at vanishing frequencies with the prefactor $61/30 \approx 2.033$. We find that the spectral tail indeed approaches this value independent of the feedback from the tail itself—provided the spectral function is computed consistently via \labelcref{eq:f_h_extraction}.

The non-trivial denominator, which was effectively missing in \cite{Fehre:2021eob}, turns out to be crucial for a consistent treatment of the spectral flow. The change induced by the feedback of the scattering tail is rather small, as shown in \Cref{fig:grav_convergence}. At intermediate scales, the additional logarithmic corrections lead to a faster decay of the spectral function.

Finally, we analyse the real and imaginary parts of the graviton propagator, which are shown in \Cref{fig:3dgravprop}. They are obtained from the KL representation for the graviton propagator, \labelcref{eq:KL}, where we insert the normalised spectral function $\rho_h^{(\textrm{ph})}$ defined in \labelcref{eq:hrhoPhys}.

In the UV, contributions to the propagator from the scattering tail fall off as $1/p^2$, similar to the behaviour of the pole. A substantial portion of the spectral weight is therefore stored in values/mass-scales around the Planck mass. For energies much larger than this scale, the loop contributions do not probe the fine details of the spectral tail, but are dominated by the spectral weight accumulated at lower mass scales. This explains why the fully converged spectral tail also follows \labelcref{eq:gravspecdecay} in the UV.

\section{Conclusion}
\label{sec:conclusion}

In the present work, we have computed self-consistent results for the graviton spectral function within on-shell renormalisation. Here, self-consistency refers to the full feedback of the spectral function into the loops, including the contributions of the scattering tail. Specifically, this guarantees the exact IR scaling of the propagator, see~\labelcref{eq:universalIR}.

On-shell renormalisation is a physical renormalisation scheme that arranges for a classical on-shell dispersion in each RG step, see \Cref{sec:OnShellRenorm}. This can be understood as a momentum-dependent rescaling of the fields, which guarantees canonical commutation relations as well as a finite total spectral weight. Within this RG scheme we are led to a normalisable positive spectral function, see \Cref{fig:GravResults}: The latter originates in the decay behaviour of the scattering tail with $1/ (\lambda^2 (\log \lambda^2)^3)$ for large spectral values $\lambda\to \infty$, see \labelcref{eq:gravspecdecay}. The fluctuation graviton within on-shell renormalisation shows important properties of a quantum field that defines an asymptotic state, despite not defining one. In particular, the appropriately normalised graviton spectral function \labelcref{eq:hrhoPhys} satisfies the spectral sum rule \labelcref{eq:Sumrulerhohphys}: the total spectral weight of the graviton is unity. It is suggestive that the respective vertices have similar physical properties, which will be discussed in a forthcoming work \cite{APRW}.

In summary, spectral flows with on-shell renormalisation allow for a formulation of asymptotically safe Lorentzian gravity in terms of RG-invariant physical building blocks that facilitate the physics interpretation of propagators and scattering couplings. In particular, this will find natural applications for the computation of asymptotically safe cross-sections, and we hope to report on respective results soon.

\section*{Acknowledgements}
We thank Gabriel Assant, Konrad Kockler, Daniel Litim, and the members of the fQCD collaboration \cite{fQCD} for discussions and work on related subjects. This work is funded by the Deutsche Forschungsgemeinschaft (DFG, German Research Foundation) under Germany’s Excellence Strategy EXC 2181/1 - 390900948 (the Heidelberg STRUCTURES Excellence Cluster) and the Collaborative Research Centre SFB 1225 - 273811115 (ISOQUANT), as well as by the Science and Technology Facilities Council under the Consolidated Grant ST/X000796/1 and the Ernest Rutherford Fellowship ST/Z510282/1.

\appendix

\section{Gauge-fixing and ghost action}
\label{app:GaugeFixing}
We are using a de-Donder type gauge-fixing,
\begin{align}
	\label{eq:Sgf}
	S_\text{gf}[\bar g,h] & = \frac1{2\alpha}\int \! \mathrm d^4 x\,
	\sqrt{\bar g}\,\bar g^{\mu\nu}F_\mu F_\nu \,
\end{align}
with
\begin{align}
	F_\mu & = \bar \nabla^\nu h_{\mu\nu} - \frac{1+\beta}4 \bar \nabla_\mu
	{h^\nu}_\nu \,.
\end{align}
Throughout this work, we use the harmonic gauge $\alpha=\beta=1$. The respective ghost action is given by
\begin{align}
	S_\text{gh}[\bar g, h, \bar c, c] & = \int \! \mathrm  d^4 x
	\,\sqrt{\bar{g}}\, \bar{c}^\mu \mathcal{M}_{\mu\nu} c^\nu \,.
\end{align}
The Faddeev-Popov operator $\mathcal{M}$ follows from a diffeomorphism
variation of the gauge-fixing condition \labelcref{eq:Sgf},
\begin{align}
	\mathcal{M}_{\mu\nu}               =
	\bar\nabla^\rho\left(g_{\mu\nu}\nabla_\rho+g_{\rho\nu}\nabla_\mu \right)
	\frac{1+\beta}{2} \bar{g}^{\rho \sigma} \bar{\nabla}_\mu\left( g_{\nu \rho}
	\nabla_\sigma \right) .
\end{align}
The ghost spectral function $\rho_c$ is parametrised in analogy to the graviton spectral function \labelcref{eq:rhoh-para} with the replacements $m_h\to k$, $Z_h\to Z_c$, and $f_h\to f_c$, $Z_c$ being the on-shell ghost wave-function renormalisation.

\section{Vertices and projections}
\label{app:Prop+Vertices}

The transverse-traceless tensor structure $\Pi_\text{TT}(p)$ is
given by
\begin{align}\label{eq:Pi_TT}\nonumber
	\Pi_\text{TT}^{\mu\nu \rho\sigma}(p) & = \Pi^{\mu(\rho}(p)
	\Pi^{\sigma)\nu}(p)  - \frac{1}{3} \Pi^{\mu\nu}(p) \Pi^{\rho\sigma}(p)\,,
	                                     &                                              \\[2ex]
	\Pi^{\mu\nu}(p)                      & = \eta^{\mu\nu} - \frac{p^\mu p^\nu}{p^2}\,,
\end{align}
where the parenthesis in the superscript stand for symmetrisation with respect to the indices $\rho$ and $\sigma$: $O_1^{(\rho}\,O_2^{\sigma)}=1/2 \,(O_1^{\rho}\,O_2^{\sigma}+O_1^{\sigma}\,O_2^{\rho})$. The subtraction in \labelcref{eq:Pi_TT} leads to $(\Pi_\text{TT})^\mu{}_\mu{}^\rho{}_\rho=0$, and we have $\Pi_\text{TT}^2= \Pi_\text{TT}$.

In the flow of the graviton two-point function in \labelcref{eq:FlowG2}, we use the flowing $n$-graviton vertices derived from $n$ metric-derivatives of the Einstein-Hilbert action \labelcref{eq:FlowingSEH} with a Newton constant $G_{\text{N},k}$ and a cosmological constant $\Lambda_k$. The dimensionless version $g$ of the Newton constant satisfies the flow equation \labelcref{eq:betafunction_g}.

\section{Spectral Flow of the graviton propagator}
\label{app:SecFlowProp}

In this Appendix, we present additional information on the diagrammatic expressions and the analytic forms of the spectral integration kernels in \Cref{sec:diagrams}. We further provide details on the generalised spectral kernels relevant to the regulator line in \Cref{sec:higherorderkernels}, and we analytically derive the UV behaviour of the on-shell renormalised graviton spectral function in \Cref{sec:UV-limit}.

\subsection{Analytic Expressions of the Diagrams}
\label{sec:diagrams}
For the diagrammatic expressions, we refer to \cite{Fehre:2021eob}. These read schematically as
\begin{align}
	{\textrm{Flow}}^{(hh)}_\text{tadpole}
	=                                     & \int_{\lambda}
	\rho^{(2)}_h(\lambda)      \,   D_\text{tadpole}(\lambda)        \,,                     \\[1ex]\nonumber
	{\textrm{Flow}}^{(hh)}_\text{3-point}
	=                                     & \, \int_{\lambda_1,\lambda_2}
	\rho^{(2)}_h(\lambda_1) \rho_h(\lambda_2)\,   D_\text{3-point}(\lambda_1,\lambda_2,p)\,, \\[1ex]\nonumber
	{\textrm{Flow}}^{(hh)}_\text{ghost} = & \,\int_{\lambda_1,\lambda_2}
	\rho^{(2)}_c(\lambda_1) \rho_c(\lambda_2)\,   D_\text{ghost}(\lambda_1,\lambda_2,p)\,.
	\label{eq:GravDiags}
\end{align}
with the momentum integrals
\begin{align}
	D_\text{tadpole}(\lambda)=                & \int_q  \frac{V_{\text{tadpole}}(p,
	q)}{(q^2+\lambda^2)^2 } \,,                                                     \\[1ex]
	D_\text{ghost}(\lambda_1,\lambda_2,p) =   & \int_q
	\frac{V_\text{ghost}(p, q)}{(q^2+\lambda_1^2)^2
	\left((p+q)^2+\lambda_2^2\right)} \,, \notag                                    \\[1ex] \notag
	D_\text{3-point}(\lambda_1,\lambda_2,p) = & \int
	_q \frac{ V_\text{3-point}(p, q)}{(q^2+\lambda_1^2)^2\left((p+q)^2+\lambda_2^2\right)}\,,
\end{align}
where we abbreviated the momentum integrals as ${\int_q = \int \frac{\mathrm{d}^d q}{(2\pi)^d}}$.
The RG-invariant vertex functions with $ V_i= \bar V_i$ in on-shell renormalisation combine the contractions of the vertices with the regulator derivative $\partial_t R_\text{\tiny CS}= (2-\eta_h)Z_h k^2$. They can be found in the supplementary material of \cite{Fehre:2021eob}.

We refrain from displaying the full, exceedingly long expressions of the diagrams; their respective imaginary parts after the (inverse) Wick-rotation, ie., $p^2\to -(\omega_+)^2$, can be reduced to a few manageable terms. For the ghost diagram, we use a classical approximation with $\rho_c(\lambda) = \delta(\lambda^2-k^2)$, leading to
\begin{align} \label{eq:ImGhostpol}
	  \text{Im}\,{\textrm{Flow}}^{(hh)}_\text{ghost}(k,\omega) &= \text{Im}\,D_\text{ghost}(k,k,\omega) \\[1ex] \nonumber 
	  &=  -2 g\, \frac{20 k^4-8 k^2 \omega^2+\omega^4 }{\omega  \sqrt{\omega^2-4 k^2}} \,\theta(\omega^2-4 k^2).
\end{align}
The imaginary part of the tadpole diagram vanishes, and the imaginary part of the graviton polarisation diagram can be reduced to
\begin{align}\label{eq:ImGravPol}
	\text{Im}\, & D_\text{3-point}(\lambda_1,\lambda_2,\omega) = \\[1ex] \nonumber
	            & \quad\frac{g \, N(\lambda_1,\lambda_2,\omega)\,\theta\!\left(\omega -\lambda_1 -\lambda_2\right)}{
		12 \omega^6 \xi_{\text{cut}}(\lambda_1,\lambda_2,\omega)\,
		(-\lambda_1^4 + 2 \lambda_1^2 \lambda_2^2 - \lambda_2^4 + \omega^4)}\,,
\end{align}
with the numerator function
\begin{widetext}
	\begin{align} \nonumber
		 & N(\lambda_1,\lambda_2,\omega) =  \bigg\lbrace-\lambda_1^{10} + \lambda_2^{10}  +11 \omega^{10}+ \lambda_2^8 \omega^2 + 4 \lambda_2^6 \omega^4 + 8 \lambda_2^4 \omega^6 + 23 \lambda_2^2 \omega^8 + \lambda_1^8 (5 \lambda_2^2 - 3 \omega^2)                                                              \\[2ex]\nonumber
		 & \hspace{1cm}- 2 \lambda_1^6 (5 \lambda_2^4 - 4 \lambda_2^2 \omega^2 + 4 \omega^4)+	2 \lambda_1^4 (5 \lambda_2^6 - 3 \lambda_2^4 \omega^2   + 14 \lambda_2^2 \omega^4 + 8 \omega^6 ) -	\lambda_1^2 (5 \lambda_2^8 + 24 \lambda_2^4 \omega^4 - 32 \lambda_2^2 \omega^6 + 15 \omega^8)\bigg\rbrace   \times \\[2ex]
		 & \hspace{1cm} \times \bigg\lbrace(-\lambda_1^2 + \lambda_2^2) \left[(\lambda_1^2 - \lambda_2^2 + \omega^2) -
			(-\lambda_1^2 + \lambda_2^2 + \omega^2)\right]   + \omega^2 \left[(\lambda_1^2 - \lambda_2^2 + \omega^2) +
			(-\lambda_1^2 + \lambda_2^2 + \omega^2)\right]\bigg\rbrace,
	\end{align}
\end{widetext}
and the cut function
\begin{align}
	\xi&_{\text{cut}}(\lambda_1,\lambda_2,w)=\\[1ex]\nonumber
	 &\qquad\qquad\sqrt{\left(\omega^2 - (\lambda_1 + \lambda_2)^2 \right)\left(\omega^2 - (\lambda_1 - \lambda_2)^2 \right)}\,.
\end{align}
To conclude the discussion of the diagrammatic expressions, we specify the final expression of the integrand in \labelcref{eq:FlowG2}, i.e., 
\begin{align}\label{eq:defImD}
	\text{Im}\, D(k,\omega,\lambda_1,\lambda_2) &=  \text{Im}\, D_\text{3-point}(\lambda_1,\lambda_2,\omega) \nonumber \\[1ex]
	 &\quad\,+ \text{Im}\, D_\text{ghost}(k,k,w)\,.
\end{align}

\subsection{Spectral representation of propagator squared}
\label{sec:higherorderkernels}
CS-flow diagrams for correlation functions always contain a regulator line of the form
\begin{align}
	G(q)\dot{R}_\text{\tiny CS}(q)G(q) = 2 k^2 G(q)^2 \,,
	\label{eq:Cutline}
\end{align}
which has the spectral representation,
\begin{align}
	G^2(q) & =\int_0^\infty \frac{\mathrm d\lambda \,\lambda}{\pi}   \frac{\rho^{(2)}(\lambda)}{(q^2+\lambda^2)^2} \,,
\end{align}
with the second-order spectral function
\begin{align}
	\partial_{w^2}\rho^{(2)}(w) & = 2\,\text{Im}\, G^2\left(-\imag (w + \imag 0^+)\right)\,.
	\label{eq:specrepGsquare}
\end{align}
Note that the Cauchy principal value of this integral does not exist for $q^2 < -\lambda_{\text{min}}$ if the pole of the integrand lies within the integration domain. In such cases, it is defined via the Hadamard finite part, or equivalently, through integration by parts, which brings the integral back to the conventional spectral form. By inserting the spectral representation of the propagator from \labelcref{eq:rhoh-para}, the imaginary part of $G^2$ has the decomposition,
\begin{widetext}
	\begin{align}
		\text{Im}\, G^2(q) = \bigg( & \pi \partial_{w^2} \delta(m_h^2 - w^2)- \frac{f_h(w) \Theta(w-2m_h)}{w^2 -m_h^2} +  \delta(m_h^2-w^2)\int\limits_{2m_h}^{\infty} \! \mathrm d\lambda^2 \frac{f_h(\lambda)}{\lambda^2 - m_h^2}-\int\limits_{2m_h}^{\infty} \! \frac{\mathrm d\lambda^2}{2\pi } \frac{f_h(\lambda)f_h(w)}{w^2-\lambda^2}\bigg) ,
		\label{eq:ImG2}
	\end{align}
\end{widetext}
where we have already used $Z_h=1$ in on-shell renormalisation. The first and third terms can be integrated analytically, while the second and fourth terms are handled numerically. This results in a delta peak at $w = m_h$ and a constant, non-zero term for $w > m_h$, which carries the nonsingular real part of the propagator at the pole as a prefactor. This constant is canceled exactly in the weight function as $w \to \infty$ by the integrated second term.

\subsection{UV Limit of the Spectral Tail}
\label{sec:UV-limit}

The UV behaviour of the spectral tail is governed by the UV limit of the imaginary part of the flow, in particular, the contribution from the graviton mass pole. Although the spectral tail has a sub-leading impact in the IR, its UV-contribution has the same scaling as the leading term, differing only by an overall factor. Furthermore, the ghost loop exhibits a similar structure, differing only in prefactors and signs. Thus, the UV scaling of the spectral tail can be extracted from the leading UV contribution due to the graviton peak, given by
\begin{align}
	\text{Im}\, \partial_t \Gamma^{(2)}(\omega \to \infty) \propto g\, \frac{\omega^3}{\sqrt{\omega^2 - 4k}}\, \theta(\omega - 2k)\,.
\end{align}
As noted in the main text, the flow is localised near the onset and diverges as $1/\sqrt{\omega^2 - 4k}$, which remains integrable and contributes a finite result upon $k$-integration. The integral only receives significant contributions for $k \approx \omega/2$. For $\omega > k$, the flow behaves as $\omega^2$, with $k$-dependence solely in the prefactor $g(k)$, which scales as $k^2$ in the IR and approaches a constant above the Planck scale. This implies that the integral is dominated by the range $M_\text{pl} < k < \omega/2$. Consequently, we can approximate the dimensionless Newton coupling by its fixed-point value and integrate the UV limit analytically,
\begin{align}\nonumber
	\text{Im}\, \Gamma^{(2)}(\omega \to \infty) &\propto \int_k^{\omega/2} \frac{\mathrm dk'}{k'}\, g^*\, \frac{\omega^3}{\sqrt{\omega^2 - 4k'}} \\[2ex]
	&=                                                    g^*\, \omega^2 \log(\frac{\omega + \sqrt{\omega^2 - 4k^2}}{2k}) .
\end{align}
For the real part, we use the subtracted Kramers–Kronig relation \labelcref{eq:KramersKronig} with $x_0 = 0$, $\text{Re}\, \Gamma^{(2)}(x_0) = 0$, and ${\partial_{\omega^2}\text{Re}\, \Gamma^{(2)}(x_0) = 0}$: we are interested only in the leading UV behaviour, which we expect to dominate over the $\omega^2$ term. Assuming $\omega \gg k$ with $k > 0$, we find,
\begin{align}\nonumber
	\text{Re}\, \Gamma^{(2)}(\omega \to \infty) \propto & \,\omega^4\, \text{PV} \int_{2k}^\infty \frac{\mathrm dt}{\pi} \frac{2\log(t)}{t(t^2 - \omega^2)} \\[2ex] \propto&\, \omega^2 \log^2(\omega) + \mathcal{O}\!\left(\omega^2 \log(\omega)\right)\,.
\end{align}
This allows us to deduce the UV scaling of the spectral tail. Since the real part dominates over both the imaginary part and the classical $\omega^2$ term in the UV, we obtain:
\begin{align}
	\rho_h(\lambda \to \infty) \propto \frac{\text{Im}\, \Gamma^{(2)}(\lambda)}{\left[\text{Re}\, \Gamma^{(2)}(\lambda)\right]^2} \propto \frac{1}{\lambda^2 \log^3(\lambda^2)}\,,
	\label{eq:UVDecay}
\end{align}
which leads to a normalisable spectral function. Note that a simple logarithmic decay with $1/(\lambda^2 \log(\lambda))$ would not suffice.

\subsection{Numerical implementation}
\label{app:numericaldetails}

We solve the flow equation for the graviton spectral function with a fixed-point iteration method: the flow equation is interpreted as an integral equation, where the spectral function on the right-hand side of~\labelcref{eq:ImGammaGrav} is fully known as a function of $\omega$ and $k$. This allows us to compute and integrate the diagrams using this input and then read out the new, fully $k$-dependent spectral function. We iterate this procedure until we achieve convergence.

The initial guess for the spectral function is the tree-level one, $\rho^{(0)}(\omega,k) = 2\pi \delta(\omega^2 - k^2)$. The on-shell renormalisation conditions \labelcref{eq:OnshellConds} ensure that the delta part of the graviton spectral function remains unchanged upon iteration, which facilitates the numerical solution. All numerical routines are implemented in \texttt{Julia}.

The flow of the spectral diagrams is computed by evaluating only the imaginary parts of the diagrams. This significantly simplifies the spectral integrals and avoids precision issues in the numerical integration. Schematically, this reads,
\begin{align}
	\text{Im}\, \Gamma^{(2)}_{i+1}(w)= \int_{k}^{k_\text{max}}\frac{\mathrm dk}{k}\,\text{Im}\,
	\text{Flow}[\rho^{(i)}_h](\omega,k)\,.
\end{align}
The real part of the integrated diagrams is then computed using a subtracted Kramers-Kronig relation,
\begin{align}\label{eq:KramersKronig}\nonumber
	 & \text{Re} \left[f(\omega)-f(\omega_0)  - (\omega^2-\omega_0^2)\,\partial_{\omega^2} f(\omega_0)\right]                                 \\[2ex]
	 & \qquad=\frac{2}{\pi}\, \text{PV} \int_{0}^\infty \mathrm dt \,\frac{(\omega^2-\omega_0^2)^2}{(w^2-\omega_0^2)^2}
	\frac{t\, \text{Im}\, f(t)}{t^2-\omega^2}\,,
\end{align}
where the subtracted Taylor expansion serves two purposes: (1) it renders the Kramers-Kronig integral finite, and (2) it allows for a convenient implementation of the on-shell renormalisation conditions \labelcref{eq:onshell-conditiondG}.

The tail of the spectral function is constructed as,
\begin{align}\nonumber
	f^{(i+1)}_h(\omega,k) = -2\, \frac{\text{Im}\, \Gamma^{(2)}_{i+1}(\omega)}{\left(\text{Im}\, \Gamma^{(2)}_{i+1}(\omega)\right)^2 + \left(\text{Re}\, \Gamma^{(2)}_{i+1}(\omega,k)\right)^2}\,,
\end{align}
where $\text{Re}\, \Gamma^{(2)}_{i+1}(\omega,k)$ also contains the term $k^2 - \omega^2$, with $-\omega^2$ originating from the frequency dependence of the classical initial condition, and $k^2$ representing the regulator. This procedure is iterated until numerical convergence is reached, i.e., when $\rho^{(i+1)}_h = \rho^{(i)}_h$ within the desired numerical precision.

The numerical implementation of one-dimensional integrals, such as the Kramers–Kronig integrals or the respective contributions of higher scatterings, uses an $h$-adaptive Gauss–Kronrod quadrature rule, implemented in \texttt{QuadGK.jl} \cite{quadgk}. We specify a relative tolerance of $10^{-10}$ with a fixed integration order of 20.  For two-dimensional spectral integrals, we use the respective multidimensional extension implemented in \texttt{HCubature.jl} \cite{HCubature,Genz1980} with a relative tolerance of $10^{-4}$ as their contributions are typically sub-leading. The absolute tolerance is set to zero in all cases.

\bibliographystyle{apsrev4-1}
\bibliography{GravityStatus}

\begin{thebibliography}{69}%
\makeatletter
\providecommand \@ifxundefined [1]{%
 \@ifx{#1\undefined}
}%
\providecommand \@ifnum [1]{%
 \ifnum #1\expandafter \@firstoftwo
 \else \expandafter \@secondoftwo
 \fi
}%
\providecommand \@ifx [1]{%
 \ifx #1\expandafter \@firstoftwo
 \else \expandafter \@secondoftwo
 \fi
}%
\providecommand \natexlab [1]{#1}%
\providecommand \enquote  [1]{``#1''}%
\providecommand \bibnamefont  [1]{#1}%
\providecommand \bibfnamefont [1]{#1}%
\providecommand \citenamefont [1]{#1}%
\providecommand \href@noop [0]{\@secondoftwo}%
\providecommand \href [0]{\begingroup \@sanitize@url \@href}%
\providecommand \@href[1]{\@@startlink{#1}\@@href}%
\providecommand \@@href[1]{\endgroup#1\@@endlink}%
\providecommand \@sanitize@url [0]{\catcode `\\12\catcode `\$12\catcode `\&12\catcode `\#12\catcode `\^12\catcode `\_12\catcode `\%12\relax}%
\providecommand \@@startlink[1]{}%
\providecommand \@@endlink[0]{}%
\providecommand \url  [0]{\begingroup\@sanitize@url \@url }%
\providecommand \@url [1]{\endgroup\@href {#1}{\urlprefix }}%
\providecommand \urlprefix  [0]{URL }%
\providecommand \Eprint [0]{\href }%
\providecommand \doibase [0]{http://dx.doi.org/}%
\providecommand \selectlanguage [0]{\@gobble}%
\providecommand \bibinfo  [0]{\@secondoftwo}%
\providecommand \bibfield  [0]{\@secondoftwo}%
\providecommand \translation [1]{[#1]}%
\providecommand \BibitemOpen [0]{}%
\providecommand \bibitemStop [0]{}%
\providecommand \bibitemNoStop [0]{.\EOS\space}%
\providecommand \EOS [0]{\spacefactor3000\relax}%
\providecommand \BibitemShut  [1]{\csname bibitem#1\endcsname}%
\let\auto@bib@innerbib\@empty
\bibitem [{\citenamefont {Fehre}\ \emph {et~al.}(2023)\citenamefont {Fehre}, \citenamefont {Litim}, \citenamefont {Pawlowski},\ and\ \citenamefont {Reichert}}]{Fehre:2021eob}%
  \BibitemOpen
  \bibfield  {author} {\bibinfo {author} {\bibfnamefont {J.}~\bibnamefont {Fehre}}, \bibinfo {author} {\bibfnamefont {D.~F.}\ \bibnamefont {Litim}}, \bibinfo {author} {\bibfnamefont {J.~M.}\ \bibnamefont {Pawlowski}}, \ and\ \bibinfo {author} {\bibfnamefont {M.}~\bibnamefont {Reichert}},\ }\href {\doibase 10.1103/PhysRevLett.130.081501} {\bibfield  {journal} {\bibinfo  {journal} {Phys. Rev. Lett.}\ }\textbf {\bibinfo {volume} {130}},\ \bibinfo {pages} {081501} (\bibinfo {year} {2023})},\ \Eprint {http://arxiv.org/abs/2111.13232} {arXiv:2111.13232 [hep-th]} \BibitemShut {NoStop}%
\bibitem [{\citenamefont {Weinberg}(1980)}]{Weinberg:1980gg}%
  \BibitemOpen
  \bibfield  {author} {\bibinfo {author} {\bibfnamefont {S.}~\bibnamefont {Weinberg}},\ }\enquote {\bibinfo {title} {{Ultraviolet divergences in quantum theories of gravitation}},}\ in\ \href@noop {} {\emph {\bibinfo {booktitle} {{General Relativity}: {An Einstein Centenary Survey}}}}\ (\bibinfo {year} {1980})\ pp.\ \bibinfo {pages} {790--831}\BibitemShut {NoStop}%
\bibitem [{\citenamefont {Reuter}(1998)}]{Reuter:1996cp}%
  \BibitemOpen
  \bibfield  {author} {\bibinfo {author} {\bibfnamefont {M.}~\bibnamefont {Reuter}},\ }\href {\doibase 10.1103/PhysRevD.57.971} {\bibfield  {journal} {\bibinfo  {journal} {Phys. Rev. D}\ }\textbf {\bibinfo {volume} {57}},\ \bibinfo {pages} {971} (\bibinfo {year} {1998})},\ \Eprint {http://arxiv.org/abs/hep-th/9605030} {arXiv:hep-th/9605030} \BibitemShut {NoStop}%
\bibitem [{\citenamefont {Bonanno}\ \emph {et~al.}(2020)\citenamefont {Bonanno}, \citenamefont {Eichhorn}, \citenamefont {Gies}, \citenamefont {Pawlowski}, \citenamefont {Percacci}, \citenamefont {Reuter}, \citenamefont {Saueressig},\ and\ \citenamefont {Vacca}}]{Bonanno:2020bil}%
  \BibitemOpen
  \bibfield  {author} {\bibinfo {author} {\bibfnamefont {A.}~\bibnamefont {Bonanno}}, \bibinfo {author} {\bibfnamefont {A.}~\bibnamefont {Eichhorn}}, \bibinfo {author} {\bibfnamefont {H.}~\bibnamefont {Gies}}, \bibinfo {author} {\bibfnamefont {J.~M.}\ \bibnamefont {Pawlowski}}, \bibinfo {author} {\bibfnamefont {R.}~\bibnamefont {Percacci}}, \bibinfo {author} {\bibfnamefont {M.}~\bibnamefont {Reuter}}, \bibinfo {author} {\bibfnamefont {F.}~\bibnamefont {Saueressig}}, \ and\ \bibinfo {author} {\bibfnamefont {G.~P.}\ \bibnamefont {Vacca}},\ }\href {\doibase 10.3389/fphy.2020.00269} {\bibfield  {journal} {\bibinfo  {journal} {Front. in Phys.}\ }\textbf {\bibinfo {volume} {8}},\ \bibinfo {pages} {269} (\bibinfo {year} {2020})},\ \Eprint {http://arxiv.org/abs/2004.06810} {arXiv:2004.06810 [gr-qc]} \BibitemShut {NoStop}%
\bibitem [{\citenamefont {Dupuis}\ \emph {et~al.}(2021)\citenamefont {Dupuis}, \citenamefont {Canet}, \citenamefont {Eichhorn}, \citenamefont {Metzner}, \citenamefont {Pawlowski}, \citenamefont {Tissier},\ and\ \citenamefont {Wschebor}}]{Dupuis:2020fhh}%
  \BibitemOpen
  \bibfield  {author} {\bibinfo {author} {\bibfnamefont {N.}~\bibnamefont {Dupuis}}, \bibinfo {author} {\bibfnamefont {L.}~\bibnamefont {Canet}}, \bibinfo {author} {\bibfnamefont {A.}~\bibnamefont {Eichhorn}}, \bibinfo {author} {\bibfnamefont {W.}~\bibnamefont {Metzner}}, \bibinfo {author} {\bibfnamefont {J.~M.}\ \bibnamefont {Pawlowski}}, \bibinfo {author} {\bibfnamefont {M.}~\bibnamefont {Tissier}}, \ and\ \bibinfo {author} {\bibfnamefont {N.}~\bibnamefont {Wschebor}},\ }\href {\doibase 10.1016/j.physrep.2021.01.001} {\bibfield  {journal} {\bibinfo  {journal} {Phys. Rept.}\ }\textbf {\bibinfo {volume} {910}},\ \bibinfo {pages} {1} (\bibinfo {year} {2021})},\ \Eprint {http://arxiv.org/abs/2006.04853} {arXiv:2006.04853 [cond-mat.stat-mech]} \BibitemShut {NoStop}%
\bibitem [{\citenamefont {Pawlowski}\ and\ \citenamefont {Reichert}(2021)}]{Pawlowski:2020qer}%
  \BibitemOpen
  \bibfield  {author} {\bibinfo {author} {\bibfnamefont {J.~M.}\ \bibnamefont {Pawlowski}}\ and\ \bibinfo {author} {\bibfnamefont {M.}~\bibnamefont {Reichert}},\ }\href {\doibase 10.3389/fphy.2020.551848} {\bibfield  {journal} {\bibinfo  {journal} {Front. in Phys.}\ }\textbf {\bibinfo {volume} {8}},\ \bibinfo {pages} {551848} (\bibinfo {year} {2021})},\ \Eprint {http://arxiv.org/abs/2007.10353} {arXiv:2007.10353 [hep-th]} \BibitemShut {NoStop}%
\bibitem [{\citenamefont {Knorr}\ \emph {et~al.}(2024)\citenamefont {Knorr}, \citenamefont {Ripken},\ and\ \citenamefont {Saueressig}}]{Knorr:2022dsx}%
  \BibitemOpen
  \bibfield  {author} {\bibinfo {author} {\bibfnamefont {B.}~\bibnamefont {Knorr}}, \bibinfo {author} {\bibfnamefont {C.}~\bibnamefont {Ripken}}, \ and\ \bibinfo {author} {\bibfnamefont {F.}~\bibnamefont {Saueressig}},\ }in\ \href {\doibase 10.1007/978-981-19-3079-9_21-1} {\emph {\bibinfo {booktitle} {Handbook of Quantum Gravity}}}\ (\bibinfo  {publisher} {Springer Nature Singapore},\ \bibinfo {address} {Singapore},\ \bibinfo {year} {2024})\ \Eprint {http://arxiv.org/abs/2210.16072} {arXiv:2210.16072 [hep-th]} \BibitemShut {NoStop}%
\bibitem [{\citenamefont {Eichhorn}\ and\ \citenamefont {Schiffer}(2022)}]{Eichhorn:2022gku}%
  \BibitemOpen
  \bibfield  {author} {\bibinfo {author} {\bibfnamefont {A.}~\bibnamefont {Eichhorn}}\ and\ \bibinfo {author} {\bibfnamefont {M.}~\bibnamefont {Schiffer}},\ }in\ \href {\doibase 10.1007/978-981-99-7681-2_22} {\emph {\bibinfo {booktitle} {Handbook of Quantum Gravity}}}\ (\bibinfo  {publisher} {Springer Nature Singapore},\ \bibinfo {address} {Singapore},\ \bibinfo {year} {2022})\ \Eprint {http://arxiv.org/abs/2212.07456} {arXiv:2212.07456 [hep-th]} \BibitemShut {NoStop}%
\bibitem [{\citenamefont {Morris}\ and\ \citenamefont {Stulga}(2023)}]{Morris:2022btf}%
  \BibitemOpen
  \bibfield  {author} {\bibinfo {author} {\bibfnamefont {T.~R.}\ \bibnamefont {Morris}}\ and\ \bibinfo {author} {\bibfnamefont {D.}~\bibnamefont {Stulga}},\ }in\ \href {\doibase 10.1007/978-981-19-3079-9_19-1} {\emph {\bibinfo {booktitle} {Handbook of Quantum Gravity}}}\ (\bibinfo  {publisher} {Springer Nature Singapore},\ \bibinfo {address} {Singapore},\ \bibinfo {year} {2023})\ pp.\ \bibinfo {pages} {1--33},\ \Eprint {http://arxiv.org/abs/2210.11356} {arXiv:2210.11356 [hep-th]} \BibitemShut {NoStop}%
\bibitem [{\citenamefont {Wetterich}(2024)}]{Wetterich:2022ncl}%
  \BibitemOpen
  \bibfield  {author} {\bibinfo {author} {\bibfnamefont {C.}~\bibnamefont {Wetterich}},\ }in\ \href {\doibase 10.1007/978-981-19-3079-9_26-1} {\emph {\bibinfo {booktitle} {Handbook of Quantum Gravity}}}\ (\bibinfo  {publisher} {Springer Nature Singapore},\ \bibinfo {address} {Singapore},\ \bibinfo {year} {2024})\ pp.\ \bibinfo {pages} {1143--1210},\ \Eprint {http://arxiv.org/abs/2211.03596} {arXiv:2211.03596 [gr-qc]} \BibitemShut {NoStop}%
\bibitem [{\citenamefont {Martini}\ \emph {et~al.}(2024)\citenamefont {Martini}, \citenamefont {Vacca},\ and\ \citenamefont {Zanusso}}]{Martini:2022sll}%
  \BibitemOpen
  \bibfield  {author} {\bibinfo {author} {\bibfnamefont {R.}~\bibnamefont {Martini}}, \bibinfo {author} {\bibfnamefont {G.~P.}\ \bibnamefont {Vacca}}, \ and\ \bibinfo {author} {\bibfnamefont {O.}~\bibnamefont {Zanusso}},\ }in\ \href {\doibase 10.1007/978-981-19-3079-9_25-1} {\emph {\bibinfo {booktitle} {Handbook of Quantum Gravity}}}\ (\bibinfo  {publisher} {Springer Nature Singapore},\ \bibinfo {address} {Singapore},\ \bibinfo {year} {2024})\ pp.\ \bibinfo {pages} {1097--1142},\ \Eprint {http://arxiv.org/abs/2210.13910} {arXiv:2210.13910 [hep-th]} \BibitemShut {NoStop}%
\bibitem [{\citenamefont {Saueressig}(2024)}]{Saueressig:2023irs}%
  \BibitemOpen
  \bibfield  {author} {\bibinfo {author} {\bibfnamefont {F.}~\bibnamefont {Saueressig}},\ }in\ \href {\doibase 10.1007/978-981-19-3079-9_16-1} {\emph {\bibinfo {booktitle} {Handbook of Quantum Gravity}}}\ (\bibinfo  {publisher} {Springer Nature Singapore},\ \bibinfo {address} {Singapore},\ \bibinfo {year} {2024})\ pp.\ \bibinfo {pages} {717--760},\ \Eprint {http://arxiv.org/abs/2302.14152} {arXiv:2302.14152 [hep-th]} \BibitemShut {NoStop}%
\bibitem [{\citenamefont {Pawlowski}\ and\ \citenamefont {Reichert}(2023)}]{Pawlowski:2023gym}%
  \BibitemOpen
  \bibfield  {author} {\bibinfo {author} {\bibfnamefont {J.~M.}\ \bibnamefont {Pawlowski}}\ and\ \bibinfo {author} {\bibfnamefont {M.}~\bibnamefont {Reichert}},\ }in\ \href {\doibase 10.1007/978-981-19-3079-9_17-1} {\emph {\bibinfo {booktitle} {Handbook of Quantum Gravity}}}\ (\bibinfo  {publisher} {Springer Nature Singapore},\ \bibinfo {address} {Singapore},\ \bibinfo {year} {2023})\ \Eprint {http://arxiv.org/abs/2309.10785} {arXiv:2309.10785 [hep-th]} \BibitemShut {NoStop}%
\bibitem [{\citenamefont {Platania}(2023)}]{Platania:2023srt}%
  \BibitemOpen
  \bibfield  {author} {\bibinfo {author} {\bibfnamefont {A.}~\bibnamefont {Platania}},\ }in\ \href {\doibase 10.1007/978-981-19-3079-9_24-1} {\emph {\bibinfo {booktitle} {Handbook of Quantum Gravity}}}\ (\bibinfo  {publisher} {Springer Nature Singapore},\ \bibinfo {address} {Singapore},\ \bibinfo {year} {2023})\ \Eprint {http://arxiv.org/abs/2302.04272} {arXiv:2302.04272 [gr-qc]} \BibitemShut {NoStop}%
\bibitem [{\citenamefont {Bonanno}(2024)}]{Bonanno:2024xne}%
  \BibitemOpen
  \bibfield  {author} {\bibinfo {author} {\bibfnamefont {A.}~\bibnamefont {Bonanno}},\ }in\ \href {\doibase 10.1007/978-981-19-3079-9_23-1} {\emph {\bibinfo {booktitle} {Handbook of Quantum Gravity}}}\ (\bibinfo  {publisher} {Springer Nature Singapore},\ \bibinfo {address} {Singapore},\ \bibinfo {year} {2024})\ pp.\ \bibinfo {pages} {1003--1029}\BibitemShut {NoStop}%
\bibitem [{\citenamefont {Reichert}(2020)}]{Reichert:2020mja}%
  \BibitemOpen
  \bibfield  {author} {\bibinfo {author} {\bibfnamefont {M.}~\bibnamefont {Reichert}},\ }\href {\doibase 10.22323/1.384.0005} {\bibfield  {journal} {\bibinfo  {journal} {PoS}\ }\textbf {\bibinfo {volume} {384}},\ \bibinfo {pages} {005} (\bibinfo {year} {2020})}\BibitemShut {NoStop}%
\bibitem [{\citenamefont {Bonanno}\ \emph {et~al.}(2022)\citenamefont {Bonanno}, \citenamefont {Denz}, \citenamefont {Pawlowski},\ and\ \citenamefont {Reichert}}]{Bonanno:2021squ}%
  \BibitemOpen
  \bibfield  {author} {\bibinfo {author} {\bibfnamefont {A.}~\bibnamefont {Bonanno}}, \bibinfo {author} {\bibfnamefont {T.}~\bibnamefont {Denz}}, \bibinfo {author} {\bibfnamefont {J.~M.}\ \bibnamefont {Pawlowski}}, \ and\ \bibinfo {author} {\bibfnamefont {M.}~\bibnamefont {Reichert}},\ }\href {\doibase 10.21468/SciPostPhys.12.1.001} {\bibfield  {journal} {\bibinfo  {journal} {SciPost Phys.}\ }\textbf {\bibinfo {volume} {12}},\ \bibinfo {pages} {001} (\bibinfo {year} {2022})},\ \Eprint {http://arxiv.org/abs/2102.02217} {arXiv:2102.02217 [hep-th]} \BibitemShut {NoStop}%
\bibitem [{\citenamefont {Kher}\ \emph {et~al.}(2025)\citenamefont {Kher}, \citenamefont {King}, \citenamefont {Litim},\ and\ \citenamefont {Reichert}}]{Kher:2025rve}%
  \BibitemOpen
  \bibfield  {author} {\bibinfo {author} {\bibfnamefont {V.}~\bibnamefont {Kher}}, \bibinfo {author} {\bibfnamefont {B.}~\bibnamefont {King}}, \bibinfo {author} {\bibfnamefont {D.~F.}\ \bibnamefont {Litim}}, \ and\ \bibinfo {author} {\bibfnamefont {M.}~\bibnamefont {Reichert}},\ }\href@noop {} {\  (\bibinfo {year} {2025})},\ \Eprint {http://arxiv.org/abs/2507.17862} {arXiv:2507.17862 [hep-th]} \BibitemShut {NoStop}%
\bibitem [{\citenamefont {Baldazzi}\ \emph {et~al.}(2019{\natexlab{a}})\citenamefont {Baldazzi}, \citenamefont {Percacci},\ and\ \citenamefont {Skrinjar}}]{Baldazzi:2018mtl}%
  \BibitemOpen
  \bibfield  {author} {\bibinfo {author} {\bibfnamefont {A.}~\bibnamefont {Baldazzi}}, \bibinfo {author} {\bibfnamefont {R.}~\bibnamefont {Percacci}}, \ and\ \bibinfo {author} {\bibfnamefont {V.}~\bibnamefont {Skrinjar}},\ }\href {\doibase 10.1088/1361-6382/ab187d} {\bibfield  {journal} {\bibinfo  {journal} {Class. Quant. Grav.}\ }\textbf {\bibinfo {volume} {36}},\ \bibinfo {pages} {105008} (\bibinfo {year} {2019}{\natexlab{a}})},\ \Eprint {http://arxiv.org/abs/1811.03369} {arXiv:1811.03369 [gr-qc]} \BibitemShut {NoStop}%
\bibitem [{\citenamefont {Baldazzi}\ \emph {et~al.}(2019{\natexlab{b}})\citenamefont {Baldazzi}, \citenamefont {Percacci},\ and\ \citenamefont {Skrinjar}}]{Baldazzi:2019kim}%
  \BibitemOpen
  \bibfield  {author} {\bibinfo {author} {\bibfnamefont {A.}~\bibnamefont {Baldazzi}}, \bibinfo {author} {\bibfnamefont {R.}~\bibnamefont {Percacci}}, \ and\ \bibinfo {author} {\bibfnamefont {V.}~\bibnamefont {Skrinjar}},\ }\href {\doibase 10.3390/sym11030373} {\bibfield  {journal} {\bibinfo  {journal} {Symmetry}\ }\textbf {\bibinfo {volume} {11}},\ \bibinfo {pages} {373} (\bibinfo {year} {2019}{\natexlab{b}})},\ \Eprint {http://arxiv.org/abs/1901.01891} {arXiv:1901.01891 [gr-qc]} \BibitemShut {NoStop}%
\bibitem [{\citenamefont {D'Angelo}\ \emph {et~al.}(2024)\citenamefont {D'Angelo}, \citenamefont {Drago}, \citenamefont {Pinamonti},\ and\ \citenamefont {Rejzner}}]{DAngelo:2022vsh}%
  \BibitemOpen
  \bibfield  {author} {\bibinfo {author} {\bibfnamefont {E.}~\bibnamefont {D'Angelo}}, \bibinfo {author} {\bibfnamefont {N.}~\bibnamefont {Drago}}, \bibinfo {author} {\bibfnamefont {N.}~\bibnamefont {Pinamonti}}, \ and\ \bibinfo {author} {\bibfnamefont {K.}~\bibnamefont {Rejzner}},\ }\href {\doibase 10.1007/s00023-023-01348-4} {\bibfield  {journal} {\bibinfo  {journal} {Annales Henri Poincare}\ }\textbf {\bibinfo {volume} {25}},\ \bibinfo {pages} {2295} (\bibinfo {year} {2024})},\ \Eprint {http://arxiv.org/abs/2202.07580} {arXiv:2202.07580 [math-ph]} \BibitemShut {NoStop}%
\bibitem [{\citenamefont {Banerjee}\ and\ \citenamefont {Niedermaier}(2022)}]{Banerjee:2022xvi}%
  \BibitemOpen
  \bibfield  {author} {\bibinfo {author} {\bibfnamefont {R.}~\bibnamefont {Banerjee}}\ and\ \bibinfo {author} {\bibfnamefont {M.}~\bibnamefont {Niedermaier}},\ }\href {\doibase 10.1016/j.nuclphysb.2022.115814} {\bibfield  {journal} {\bibinfo  {journal} {Nucl. Phys. B}\ }\textbf {\bibinfo {volume} {980}},\ \bibinfo {pages} {115814} (\bibinfo {year} {2022})},\ \Eprint {http://arxiv.org/abs/2201.02575} {arXiv:2201.02575 [hep-th]} \BibitemShut {NoStop}%
\bibitem [{\citenamefont {D'Angelo}\ and\ \citenamefont {Rejzner}(2025)}]{DAngelo:2023tis}%
  \BibitemOpen
  \bibfield  {author} {\bibinfo {author} {\bibfnamefont {E.}~\bibnamefont {D'Angelo}}\ and\ \bibinfo {author} {\bibfnamefont {K.}~\bibnamefont {Rejzner}},\ }\href {\doibase 10.1007/s00023-024-01535-x} {\bibfield  {journal} {\bibinfo  {journal} {Annales Henri Poincaré}\ ,\ \bibinfo {pages} {1424}} (\bibinfo {year} {2025})},\ \Eprint {http://arxiv.org/abs/2303.01479} {arXiv:2303.01479 [math-ph]} \BibitemShut {NoStop}%
\bibitem [{\citenamefont {D'Angelo}(2024)}]{DAngelo:2023wje}%
  \BibitemOpen
  \bibfield  {author} {\bibinfo {author} {\bibfnamefont {E.}~\bibnamefont {D'Angelo}},\ }\href {\doibase 10.1103/PhysRevD.109.066012} {\bibfield  {journal} {\bibinfo  {journal} {Phys. Rev. D}\ }\textbf {\bibinfo {volume} {109}},\ \bibinfo {pages} {066012} (\bibinfo {year} {2024})},\ \Eprint {http://arxiv.org/abs/2310.20603} {arXiv:2310.20603 [hep-th]} \BibitemShut {NoStop}%
\bibitem [{\citenamefont {Banerjee}\ and\ \citenamefont {Niedermaier}(2025)}]{Banerjee:2024tap}%
  \BibitemOpen
  \bibfield  {author} {\bibinfo {author} {\bibfnamefont {R.}~\bibnamefont {Banerjee}}\ and\ \bibinfo {author} {\bibfnamefont {M.}~\bibnamefont {Niedermaier}},\ }\href {\doibase 10.1088/1361-6382/adc9ef} {\bibfield  {journal} {\bibinfo  {journal} {Class. Quant. Grav.}\ }\textbf {\bibinfo {volume} {42}},\ \bibinfo {pages} {095003} (\bibinfo {year} {2025})},\ \Eprint {http://arxiv.org/abs/2406.06047} {arXiv:2406.06047 [math-ph]} \BibitemShut {NoStop}%
\bibitem [{\citenamefont {Thiemann}(2024)}]{Thiemann:2024vjx}%
  \BibitemOpen
  \bibfield  {author} {\bibinfo {author} {\bibfnamefont {T.}~\bibnamefont {Thiemann}},\ }\href {\doibase 10.1007/JHEP10(2024)013} {\bibfield  {journal} {\bibinfo  {journal} {JHEP}\ }\textbf {\bibinfo {volume} {10}},\ \bibinfo {pages} {013} (\bibinfo {year} {2024})},\ \Eprint {http://arxiv.org/abs/2404.18220} {arXiv:2404.18220 [hep-th]} \BibitemShut {NoStop}%
\bibitem [{\citenamefont {Ferrero}\ and\ \citenamefont {Thiemann}(2024)}]{Ferrero:2024rvi}%
  \BibitemOpen
  \bibfield  {author} {\bibinfo {author} {\bibfnamefont {R.}~\bibnamefont {Ferrero}}\ and\ \bibinfo {author} {\bibfnamefont {T.}~\bibnamefont {Thiemann}},\ }\href {\doibase 10.3390/universe10110410} {\bibfield  {journal} {\bibinfo  {journal} {Universe}\ }\textbf {\bibinfo {volume} {10}},\ \bibinfo {pages} {410} (\bibinfo {year} {2024})},\ \Eprint {http://arxiv.org/abs/2404.18224} {arXiv:2404.18224 [hep-th]} \BibitemShut {NoStop}%
\bibitem [{\citenamefont {D’Angelo}\ \emph {et~al.}(2025)\citenamefont {D’Angelo}, \citenamefont {Ferrero},\ and\ \citenamefont {Fröb}}]{DAngelo:2025yoy}%
  \BibitemOpen
  \bibfield  {author} {\bibinfo {author} {\bibfnamefont {E.}~\bibnamefont {D’Angelo}}, \bibinfo {author} {\bibfnamefont {R.}~\bibnamefont {Ferrero}}, \ and\ \bibinfo {author} {\bibfnamefont {M.~B.}\ \bibnamefont {Fröb}},\ }\href {\doibase 10.1088/1361-6382/ade193} {\bibfield  {journal} {\bibinfo  {journal} {Classical and Quantum Gravity}\ }\textbf {\bibinfo {volume} {42}},\ \bibinfo {pages} {125008} (\bibinfo {year} {2025})},\ \Eprint {http://arxiv.org/abs/2502.05135} {arXiv:2502.05135 [hep-th]} \BibitemShut {NoStop}%
\bibitem [{\citenamefont {Knorr}(2019)}]{Knorr:2018fdu}%
  \BibitemOpen
  \bibfield  {author} {\bibinfo {author} {\bibfnamefont {B.}~\bibnamefont {Knorr}},\ }\href {\doibase 10.1016/j.physletb.2019.01.070} {\bibfield  {journal} {\bibinfo  {journal} {Phys. Lett. B}\ }\textbf {\bibinfo {volume} {792}},\ \bibinfo {pages} {142} (\bibinfo {year} {2019})},\ \Eprint {http://arxiv.org/abs/1810.07971} {arXiv:1810.07971 [hep-th]} \BibitemShut {NoStop}%
\bibitem [{\citenamefont {Eichhorn}\ \emph {et~al.}(2020)\citenamefont {Eichhorn}, \citenamefont {Platania},\ and\ \citenamefont {Schiffer}}]{Eichhorn:2019ybe}%
  \BibitemOpen
  \bibfield  {author} {\bibinfo {author} {\bibfnamefont {A.}~\bibnamefont {Eichhorn}}, \bibinfo {author} {\bibfnamefont {A.}~\bibnamefont {Platania}}, \ and\ \bibinfo {author} {\bibfnamefont {M.}~\bibnamefont {Schiffer}},\ }\href {\doibase 10.1103/PhysRevD.102.026007} {\bibfield  {journal} {\bibinfo  {journal} {Phys. Rev. D}\ }\textbf {\bibinfo {volume} {102}},\ \bibinfo {pages} {026007} (\bibinfo {year} {2020})},\ \Eprint {http://arxiv.org/abs/1911.10066} {arXiv:1911.10066 [hep-th]} \BibitemShut {NoStop}%
\bibitem [{\citenamefont {Knorr}\ \emph {et~al.}(2022{\natexlab{a}})\citenamefont {Knorr}, \citenamefont {Platania},\ and\ \citenamefont {Schiffer}}]{Knorr:2022mvn}%
  \BibitemOpen
  \bibfield  {author} {\bibinfo {author} {\bibfnamefont {B.}~\bibnamefont {Knorr}}, \bibinfo {author} {\bibfnamefont {A.}~\bibnamefont {Platania}}, \ and\ \bibinfo {author} {\bibfnamefont {M.}~\bibnamefont {Schiffer}},\ }\href {\doibase 10.1103/PhysRevD.106.126002} {\bibfield  {journal} {\bibinfo  {journal} {Phys. Rev. D}\ }\textbf {\bibinfo {volume} {106}},\ \bibinfo {pages} {126002} (\bibinfo {year} {2022}{\natexlab{a}})},\ \Eprint {http://arxiv.org/abs/2205.13558} {arXiv:2205.13558 [hep-th]} \BibitemShut {NoStop}%
\bibitem [{\citenamefont {Saueressig}\ and\ \citenamefont {Wang}(2023)}]{Saueressig:2023tfy}%
  \BibitemOpen
  \bibfield  {author} {\bibinfo {author} {\bibfnamefont {F.}~\bibnamefont {Saueressig}}\ and\ \bibinfo {author} {\bibfnamefont {J.}~\bibnamefont {Wang}},\ }\href {\doibase 10.1007/JHEP09(2023)064} {\bibfield  {journal} {\bibinfo  {journal} {JHEP}\ }\textbf {\bibinfo {volume} {09}},\ \bibinfo {pages} {064} (\bibinfo {year} {2023})},\ \Eprint {http://arxiv.org/abs/2306.10408} {arXiv:2306.10408 [hep-th]} \BibitemShut {NoStop}%
\bibitem [{\citenamefont {Korver}\ \emph {et~al.}(2024)\citenamefont {Korver}, \citenamefont {Saueressig},\ and\ \citenamefont {Wang}}]{Korver:2024sam}%
  \BibitemOpen
  \bibfield  {author} {\bibinfo {author} {\bibfnamefont {G.}~\bibnamefont {Korver}}, \bibinfo {author} {\bibfnamefont {F.}~\bibnamefont {Saueressig}}, \ and\ \bibinfo {author} {\bibfnamefont {J.}~\bibnamefont {Wang}},\ }\href {\doibase 10.1016/j.physletb.2024.138789} {\bibfield  {journal} {\bibinfo  {journal} {Phys. Lett. B}\ }\textbf {\bibinfo {volume} {855}},\ \bibinfo {pages} {138789} (\bibinfo {year} {2024})},\ \Eprint {http://arxiv.org/abs/2402.01260} {arXiv:2402.01260 [hep-th]} \BibitemShut {NoStop}%
\bibitem [{\citenamefont {Saueressig}\ and\ \citenamefont {Wang}(2025)}]{Saueressig:2025ypi}%
  \BibitemOpen
  \bibfield  {author} {\bibinfo {author} {\bibfnamefont {F.}~\bibnamefont {Saueressig}}\ and\ \bibinfo {author} {\bibfnamefont {J.}~\bibnamefont {Wang}},\ }\href {\doibase 10.1103/PhysRevD.111.106007} {\bibfield  {journal} {\bibinfo  {journal} {Phys. Rev. D}\ }\textbf {\bibinfo {volume} {111}},\ \bibinfo {pages} {106007} (\bibinfo {year} {2025})},\ \Eprint {http://arxiv.org/abs/2501.03752} {arXiv:2501.03752 [hep-th]} \BibitemShut {NoStop}%
\bibitem [{\citenamefont {Platania}\ and\ \citenamefont {Wetterich}(2020)}]{Platania:2020knd}%
  \BibitemOpen
  \bibfield  {author} {\bibinfo {author} {\bibfnamefont {A.}~\bibnamefont {Platania}}\ and\ \bibinfo {author} {\bibfnamefont {C.}~\bibnamefont {Wetterich}},\ }\href {\doibase 10.1016/j.physletb.2020.135911} {\bibfield  {journal} {\bibinfo  {journal} {Phys. Lett. B}\ }\textbf {\bibinfo {volume} {811}},\ \bibinfo {pages} {135911} (\bibinfo {year} {2020})},\ \Eprint {http://arxiv.org/abs/2009.06637} {arXiv:2009.06637 [hep-th]} \BibitemShut {NoStop}%
\bibitem [{\citenamefont {Platania}(2022)}]{Platania:2022gtt}%
  \BibitemOpen
  \bibfield  {author} {\bibinfo {author} {\bibfnamefont {A.}~\bibnamefont {Platania}},\ }\href {\doibase 10.1007/JHEP09(2022)167} {\bibfield  {journal} {\bibinfo  {journal} {JHEP}\ }\textbf {\bibinfo {volume} {09}},\ \bibinfo {pages} {167} (\bibinfo {year} {2022})},\ \Eprint {http://arxiv.org/abs/2206.04072} {arXiv:2206.04072 [hep-th]} \BibitemShut {NoStop}%
\bibitem [{\citenamefont {Draper}\ \emph {et~al.}(2020)\citenamefont {Draper}, \citenamefont {Knorr}, \citenamefont {Ripken},\ and\ \citenamefont {Saueressig}}]{Draper:2020bop}%
  \BibitemOpen
  \bibfield  {author} {\bibinfo {author} {\bibfnamefont {T.}~\bibnamefont {Draper}}, \bibinfo {author} {\bibfnamefont {B.}~\bibnamefont {Knorr}}, \bibinfo {author} {\bibfnamefont {C.}~\bibnamefont {Ripken}}, \ and\ \bibinfo {author} {\bibfnamefont {F.}~\bibnamefont {Saueressig}},\ }\href {\doibase 10.1103/PhysRevLett.125.181301} {\bibfield  {journal} {\bibinfo  {journal} {Phys. Rev. Lett.}\ }\textbf {\bibinfo {volume} {125}},\ \bibinfo {pages} {181301} (\bibinfo {year} {2020})},\ \Eprint {http://arxiv.org/abs/2007.00733} {arXiv:2007.00733 [hep-th]} \BibitemShut {NoStop}%
\bibitem [{\citenamefont {Knorr}\ and\ \citenamefont {Ripken}(2021)}]{Knorr:2020bjm}%
  \BibitemOpen
  \bibfield  {author} {\bibinfo {author} {\bibfnamefont {B.}~\bibnamefont {Knorr}}\ and\ \bibinfo {author} {\bibfnamefont {C.}~\bibnamefont {Ripken}},\ }\href {\doibase 10.1103/PhysRevD.103.105019} {\bibfield  {journal} {\bibinfo  {journal} {Phys. Rev. D}\ }\textbf {\bibinfo {volume} {103}},\ \bibinfo {pages} {105019} (\bibinfo {year} {2021})},\ \Eprint {http://arxiv.org/abs/2012.05144} {arXiv:2012.05144 [hep-th]} \BibitemShut {NoStop}%
\bibitem [{\citenamefont {Knorr}\ \emph {et~al.}(2022{\natexlab{b}})\citenamefont {Knorr}, \citenamefont {Pirlo}, \citenamefont {Ripken},\ and\ \citenamefont {Saueressig}}]{Knorr:2022lzn}%
  \BibitemOpen
  \bibfield  {author} {\bibinfo {author} {\bibfnamefont {B.}~\bibnamefont {Knorr}}, \bibinfo {author} {\bibfnamefont {S.}~\bibnamefont {Pirlo}}, \bibinfo {author} {\bibfnamefont {C.}~\bibnamefont {Ripken}}, \ and\ \bibinfo {author} {\bibfnamefont {F.}~\bibnamefont {Saueressig}},\ }\href@noop {} {\  (\bibinfo {year} {2022}{\natexlab{b}})},\ \Eprint {http://arxiv.org/abs/2205.01738} {arXiv:2205.01738 [hep-th]} \BibitemShut {NoStop}%
\bibitem [{\citenamefont {Pastor-Guti\'errez}\ \emph {et~al.}(2025)\citenamefont {Pastor-Guti\'errez}, \citenamefont {Pawlowski}, \citenamefont {Reichert},\ and\ \citenamefont {Ruisi}}]{Pastor-Gutierrez:2024sbt}%
  \BibitemOpen
  \bibfield  {author} {\bibinfo {author} {\bibfnamefont {A.}~\bibnamefont {Pastor-Guti\'errez}}, \bibinfo {author} {\bibfnamefont {J.~M.}\ \bibnamefont {Pawlowski}}, \bibinfo {author} {\bibfnamefont {M.}~\bibnamefont {Reichert}}, \ and\ \bibinfo {author} {\bibfnamefont {G.}~\bibnamefont {Ruisi}},\ }\href {\doibase 10.1103/PhysRevD.111.106005} {\bibfield  {journal} {\bibinfo  {journal} {Phys. Rev. D}\ }\textbf {\bibinfo {volume} {111}},\ \bibinfo {pages} {106005} (\bibinfo {year} {2025})},\ \Eprint {http://arxiv.org/abs/2412.13800} {arXiv:2412.13800 [hep-ph]} \BibitemShut {NoStop}%
\bibitem [{\citenamefont {Knorr}\ and\ \citenamefont {Platania}(2025)}]{Knorr:2024yiu}%
  \BibitemOpen
  \bibfield  {author} {\bibinfo {author} {\bibfnamefont {B.}~\bibnamefont {Knorr}}\ and\ \bibinfo {author} {\bibfnamefont {A.}~\bibnamefont {Platania}},\ }\href {\doibase 10.1007/JHEP03(2025)003} {\bibfield  {journal} {\bibinfo  {journal} {JHEP}\ }\textbf {\bibinfo {volume} {03}},\ \bibinfo {pages} {003} (\bibinfo {year} {2025})},\ \Eprint {http://arxiv.org/abs/2405.08860} {arXiv:2405.08860 [hep-th]} \BibitemShut {NoStop}%
\bibitem [{\citenamefont {Eichhorn}\ \emph {et~al.}(2025)\citenamefont {Eichhorn}, \citenamefont {Schiffer},\ and\ \citenamefont {Pedersen}}]{Eichhorn:2024wba}%
  \BibitemOpen
  \bibfield  {author} {\bibinfo {author} {\bibfnamefont {A.}~\bibnamefont {Eichhorn}}, \bibinfo {author} {\bibfnamefont {M.}~\bibnamefont {Schiffer}}, \ and\ \bibinfo {author} {\bibfnamefont {A.~O.}\ \bibnamefont {Pedersen}},\ }\href {\doibase 10.1140/epjc/s10052-025-14449-7} {\bibfield  {journal} {\bibinfo  {journal} {Eur. Phys. J. C}\ }\textbf {\bibinfo {volume} {85}},\ \bibinfo {pages} {733} (\bibinfo {year} {2025})},\ \Eprint {http://arxiv.org/abs/2405.08862} {arXiv:2405.08862 [hep-th]} \BibitemShut {NoStop}%
\bibitem [{\citenamefont {Horak}\ \emph {et~al.}(2020)\citenamefont {Horak}, \citenamefont {Pawlowski},\ and\ \citenamefont {Wink}}]{Horak:2020eng}%
  \BibitemOpen
  \bibfield  {author} {\bibinfo {author} {\bibfnamefont {J.}~\bibnamefont {Horak}}, \bibinfo {author} {\bibfnamefont {J.~M.}\ \bibnamefont {Pawlowski}}, \ and\ \bibinfo {author} {\bibfnamefont {N.}~\bibnamefont {Wink}},\ }\href {\doibase 10.1103/PhysRevD.102.125016} {\bibfield  {journal} {\bibinfo  {journal} {Phys. Rev. D}\ }\textbf {\bibinfo {volume} {102}},\ \bibinfo {pages} {125016} (\bibinfo {year} {2020})},\ \Eprint {http://arxiv.org/abs/2006.09778} {arXiv:2006.09778 [hep-th]} \BibitemShut {NoStop}%
\bibitem [{\citenamefont {Braun}\ \emph {et~al.}(2023)\citenamefont {Braun} \emph {et~al.}}]{Braun:2022mgx}%
  \BibitemOpen
  \bibfield  {author} {\bibinfo {author} {\bibfnamefont {J.}~\bibnamefont {Braun}} \emph {et~al.},\ }\href {\doibase 10.21468/SciPostPhysCore.6.3.061} {\bibfield  {journal} {\bibinfo  {journal} {SciPost Phys. Core}\ }\textbf {\bibinfo {volume} {6}},\ \bibinfo {pages} {061} (\bibinfo {year} {2023})},\ \Eprint {http://arxiv.org/abs/2206.10232} {arXiv:2206.10232 [hep-th]} \BibitemShut {NoStop}%
\bibitem [{\citenamefont {Horak}\ \emph {et~al.}(2021)\citenamefont {Horak}, \citenamefont {Papavassiliou}, \citenamefont {Pawlowski},\ and\ \citenamefont {Wink}}]{Horak:2021pfr}%
  \BibitemOpen
  \bibfield  {author} {\bibinfo {author} {\bibfnamefont {J.}~\bibnamefont {Horak}}, \bibinfo {author} {\bibfnamefont {J.}~\bibnamefont {Papavassiliou}}, \bibinfo {author} {\bibfnamefont {J.~M.}\ \bibnamefont {Pawlowski}}, \ and\ \bibinfo {author} {\bibfnamefont {N.}~\bibnamefont {Wink}},\ }\href {\doibase 10.1103/PhysRevD.104.074017} {\bibfield  {journal} {\bibinfo  {journal} {Phys. Rev. D}\ }\textbf {\bibinfo {volume} {104}},\ \bibinfo {pages} {074017} (\bibinfo {year} {2021})},\ \Eprint {http://arxiv.org/abs/2103.16175} {arXiv:2103.16175 [hep-th]} \BibitemShut {NoStop}%
\bibitem [{\citenamefont {Horak}\ \emph {et~al.}(2025)\citenamefont {Horak}, \citenamefont {Pawlowski},\ and\ \citenamefont {Wink}}]{Horak:2022myj}%
  \BibitemOpen
  \bibfield  {author} {\bibinfo {author} {\bibfnamefont {J.}~\bibnamefont {Horak}}, \bibinfo {author} {\bibfnamefont {J.~M.}\ \bibnamefont {Pawlowski}}, \ and\ \bibinfo {author} {\bibfnamefont {N.}~\bibnamefont {Wink}},\ }\href {\doibase 10.21468/SciPostPhysCore.8.3.048} {\bibfield  {journal} {\bibinfo  {journal} {SciPost Phys. Core}\ }\textbf {\bibinfo {volume} {8}},\ \bibinfo {pages} {048} (\bibinfo {year} {2025})},\ \Eprint {http://arxiv.org/abs/2202.09333} {arXiv:2202.09333 [hep-th]} \BibitemShut {NoStop}%
\bibitem [{\citenamefont {Horak}\ \emph {et~al.}(2023)\citenamefont {Horak}, \citenamefont {Pawlowski},\ and\ \citenamefont {Wink}}]{Horak:2022aza}%
  \BibitemOpen
  \bibfield  {author} {\bibinfo {author} {\bibfnamefont {J.}~\bibnamefont {Horak}}, \bibinfo {author} {\bibfnamefont {J.~M.}\ \bibnamefont {Pawlowski}}, \ and\ \bibinfo {author} {\bibfnamefont {N.}~\bibnamefont {Wink}},\ }\href {\doibase 10.21468/SciPostPhys.15.4.149} {\bibfield  {journal} {\bibinfo  {journal} {SciPost Phys.}\ }\textbf {\bibinfo {volume} {15}},\ \bibinfo {pages} {149} (\bibinfo {year} {2023})},\ \Eprint {http://arxiv.org/abs/2210.07597} {arXiv:2210.07597 [hep-ph]} \BibitemShut {NoStop}%
\bibitem [{\citenamefont {Horak}\ \emph {et~al.}(2024)\citenamefont {Horak}, \citenamefont {Ihssen}, \citenamefont {Pawlowski}, \citenamefont {Wessely},\ and\ \citenamefont {Wink}}]{Horak:2023hkp}%
  \BibitemOpen
  \bibfield  {author} {\bibinfo {author} {\bibfnamefont {J.}~\bibnamefont {Horak}}, \bibinfo {author} {\bibfnamefont {F.}~\bibnamefont {Ihssen}}, \bibinfo {author} {\bibfnamefont {J.~M.}\ \bibnamefont {Pawlowski}}, \bibinfo {author} {\bibfnamefont {J.}~\bibnamefont {Wessely}}, \ and\ \bibinfo {author} {\bibfnamefont {N.}~\bibnamefont {Wink}},\ }\href {\doibase 10.1103/PhysRevD.110.056009} {\bibfield  {journal} {\bibinfo  {journal} {Phys. Rev. D}\ }\textbf {\bibinfo {volume} {110}},\ \bibinfo {pages} {056009} (\bibinfo {year} {2024})},\ \Eprint {http://arxiv.org/abs/2303.16719} {arXiv:2303.16719 [hep-th]} \BibitemShut {NoStop}%
\bibitem [{\citenamefont {Pawlowski}\ and\ \citenamefont {Wessely}(2024)}]{Pawlowski:2024kxc}%
  \BibitemOpen
  \bibfield  {author} {\bibinfo {author} {\bibfnamefont {J.~M.}\ \bibnamefont {Pawlowski}}\ and\ \bibinfo {author} {\bibfnamefont {J.}~\bibnamefont {Wessely}},\ }\href@noop {} {\  (\bibinfo {year} {2024})},\ \Eprint {http://arxiv.org/abs/2412.12033} {arXiv:2412.12033 [hep-ph]} \BibitemShut {NoStop}%
\bibitem [{\citenamefont {Symanzik}(1970)}]{Symanzik:1970rt}%
  \BibitemOpen
  \bibfield  {author} {\bibinfo {author} {\bibfnamefont {K.}~\bibnamefont {Symanzik}},\ }\href {\doibase 10.1007/BF01649434} {\bibfield  {journal} {\bibinfo  {journal} {Commun. Math. Phys.}\ }\textbf {\bibinfo {volume} {18}},\ \bibinfo {pages} {227} (\bibinfo {year} {1970})}\BibitemShut {NoStop}%
\bibitem [{\citenamefont {Wetterich}(1993)}]{Wetterich:1992yh}%
  \BibitemOpen
  \bibfield  {author} {\bibinfo {author} {\bibfnamefont {C.}~\bibnamefont {Wetterich}},\ }\href {\doibase 10.1016/0370-2693(93)90726-X} {\bibfield  {journal} {\bibinfo  {journal} {Phys. Lett. B}\ }\textbf {\bibinfo {volume} {301}},\ \bibinfo {pages} {90} (\bibinfo {year} {1993})},\ \Eprint {http://arxiv.org/abs/1710.05815} {arXiv:1710.05815 [hep-th]} \BibitemShut {NoStop}%
\bibitem [{\citenamefont {Pawlowski}(2007)}]{Pawlowski:2005xe}%
  \BibitemOpen
  \bibfield  {author} {\bibinfo {author} {\bibfnamefont {J.~M.}\ \bibnamefont {Pawlowski}},\ }\href {\doibase 10.1016/j.aop.2007.01.007} {\bibfield  {journal} {\bibinfo  {journal} {Annals Phys.}\ }\textbf {\bibinfo {volume} {322}},\ \bibinfo {pages} {2831} (\bibinfo {year} {2007})},\ \Eprint {http://arxiv.org/abs/hep-th/0512261} {arXiv:hep-th/0512261} \BibitemShut {NoStop}%
\bibitem [{\citenamefont {Ellwanger}(1994)}]{Ellwanger:1993mw}%
  \BibitemOpen
  \bibfield  {author} {\bibinfo {author} {\bibfnamefont {U.}~\bibnamefont {Ellwanger}},\ }\href {\doibase 10.1007/BF01555911} {\bibfield  {journal} {\bibinfo  {journal} {Z. Phys. C}\ }\textbf {\bibinfo {volume} {62}},\ \bibinfo {pages} {503} (\bibinfo {year} {1994})},\ \Eprint {http://arxiv.org/abs/hep-ph/9308260} {arXiv:hep-ph/9308260} \BibitemShut {NoStop}%
\bibitem [{\citenamefont {Morris}(1994)}]{Morris:1993qb}%
  \BibitemOpen
  \bibfield  {author} {\bibinfo {author} {\bibfnamefont {T.~R.}\ \bibnamefont {Morris}},\ }\href {\doibase 10.1142/S0217751X94000972} {\bibfield  {journal} {\bibinfo  {journal} {Int. J. Mod. Phys. A}\ }\textbf {\bibinfo {volume} {9}},\ \bibinfo {pages} {2411} (\bibinfo {year} {1994})},\ \Eprint {http://arxiv.org/abs/hep-ph/9308265} {arXiv:hep-ph/9308265} \BibitemShut {NoStop}%
\bibitem [{\citenamefont {Kockler}\ \emph {et~al.}(2025)\citenamefont {Kockler}, \citenamefont {Pawlowski},\ and\ \citenamefont {Wessely}}]{Kockler:2025kdt}%
  \BibitemOpen
  \bibfield  {author} {\bibinfo {author} {\bibfnamefont {K.}~\bibnamefont {Kockler}}, \bibinfo {author} {\bibfnamefont {J.~M.}\ \bibnamefont {Pawlowski}}, \ and\ \bibinfo {author} {\bibfnamefont {J.}~\bibnamefont {Wessely}},\ }\href@noop {} {\  (\bibinfo {year} {2025})},\ \Eprint {http://arxiv.org/abs/2506.09142} {arXiv:2506.09142 [hep-th]} \BibitemShut {NoStop}%
\bibitem [{\citenamefont {Assant}\ \emph {et~al.}(2025{\natexlab{a}})\citenamefont {Assant}, \citenamefont {Pawlowski}, \citenamefont {Reichert},\ and\ \citenamefont {Wessely}}]{APRW}%
  \BibitemOpen
  \bibfield  {author} {\bibinfo {author} {\bibfnamefont {G.}~\bibnamefont {Assant}}, \bibinfo {author} {\bibfnamefont {J.~M.}\ \bibnamefont {Pawlowski}}, \bibinfo {author} {\bibfnamefont {M.}~\bibnamefont {Reichert}}, \ and\ \bibinfo {author} {\bibfnamefont {J.}~\bibnamefont {Wessely}},\ }\href@noop {} {\  (\bibinfo {year} {2025}{\natexlab{a}})},\ \bibinfo {note} {in preparation}\BibitemShut {NoStop}%
\bibitem [{\citenamefont {Christiansen}\ \emph {et~al.}(2016)\citenamefont {Christiansen}, \citenamefont {Knorr}, \citenamefont {Pawlowski},\ and\ \citenamefont {Rodigast}}]{Christiansen:2014raa}%
  \BibitemOpen
  \bibfield  {author} {\bibinfo {author} {\bibfnamefont {N.}~\bibnamefont {Christiansen}}, \bibinfo {author} {\bibfnamefont {B.}~\bibnamefont {Knorr}}, \bibinfo {author} {\bibfnamefont {J.~M.}\ \bibnamefont {Pawlowski}}, \ and\ \bibinfo {author} {\bibfnamefont {A.}~\bibnamefont {Rodigast}},\ }\href {\doibase 10.1103/PhysRevD.93.044036} {\bibfield  {journal} {\bibinfo  {journal} {Phys. Rev. D}\ }\textbf {\bibinfo {volume} {93}},\ \bibinfo {pages} {044036} (\bibinfo {year} {2016})},\ \Eprint {http://arxiv.org/abs/1403.1232} {arXiv:1403.1232 [hep-th]} \BibitemShut {NoStop}%
\bibitem [{\citenamefont {Christiansen}\ \emph {et~al.}(2015)\citenamefont {Christiansen}, \citenamefont {Knorr}, \citenamefont {Meibohm}, \citenamefont {Pawlowski},\ and\ \citenamefont {Reichert}}]{Christiansen:2015rva}%
  \BibitemOpen
  \bibfield  {author} {\bibinfo {author} {\bibfnamefont {N.}~\bibnamefont {Christiansen}}, \bibinfo {author} {\bibfnamefont {B.}~\bibnamefont {Knorr}}, \bibinfo {author} {\bibfnamefont {J.}~\bibnamefont {Meibohm}}, \bibinfo {author} {\bibfnamefont {J.~M.}\ \bibnamefont {Pawlowski}}, \ and\ \bibinfo {author} {\bibfnamefont {M.}~\bibnamefont {Reichert}},\ }\href {\doibase 10.1103/PhysRevD.92.121501} {\bibfield  {journal} {\bibinfo  {journal} {Phys. Rev. D}\ }\textbf {\bibinfo {volume} {92}},\ \bibinfo {pages} {121501} (\bibinfo {year} {2015})},\ \Eprint {http://arxiv.org/abs/1506.07016} {arXiv:1506.07016 [hep-th]} \BibitemShut {NoStop}%
\bibitem [{\citenamefont {Denz}\ \emph {et~al.}(2018)\citenamefont {Denz}, \citenamefont {Pawlowski},\ and\ \citenamefont {Reichert}}]{Denz:2016qks}%
  \BibitemOpen
  \bibfield  {author} {\bibinfo {author} {\bibfnamefont {T.}~\bibnamefont {Denz}}, \bibinfo {author} {\bibfnamefont {J.~M.}\ \bibnamefont {Pawlowski}}, \ and\ \bibinfo {author} {\bibfnamefont {M.}~\bibnamefont {Reichert}},\ }\href {\doibase 10.1140/epjc/s10052-018-5806-0} {\bibfield  {journal} {\bibinfo  {journal} {Eur. Phys. J. C}\ }\textbf {\bibinfo {volume} {78}},\ \bibinfo {pages} {336} (\bibinfo {year} {2018})},\ \Eprint {http://arxiv.org/abs/1612.07315} {arXiv:1612.07315 [hep-th]} \BibitemShut {NoStop}%
\bibitem [{\citenamefont {Ihssen}\ \emph {et~al.}(2024)\citenamefont {Ihssen}, \citenamefont {Pawlowski}, \citenamefont {Sattler},\ and\ \citenamefont {Wink}}]{Ihssen:2024miv}%
  \BibitemOpen
  \bibfield  {author} {\bibinfo {author} {\bibfnamefont {F.}~\bibnamefont {Ihssen}}, \bibinfo {author} {\bibfnamefont {J.~M.}\ \bibnamefont {Pawlowski}}, \bibinfo {author} {\bibfnamefont {F.~R.}\ \bibnamefont {Sattler}}, \ and\ \bibinfo {author} {\bibfnamefont {N.}~\bibnamefont {Wink}},\ }\href@noop {} {\  (\bibinfo {year} {2024})},\ \Eprint {http://arxiv.org/abs/2408.08413} {arXiv:2408.08413 [hep-ph]} \BibitemShut {NoStop}%
\bibitem [{\citenamefont {Assant}\ \emph {et~al.}(2025{\natexlab{b}})\citenamefont {Assant}, \citenamefont {Litim},\ and\ \citenamefont {Reichert}}]{ALR}%
  \BibitemOpen
  \bibfield  {author} {\bibinfo {author} {\bibfnamefont {G.}~\bibnamefont {Assant}}, \bibinfo {author} {\bibfnamefont {D.~F.}\ \bibnamefont {Litim}}, \ and\ \bibinfo {author} {\bibfnamefont {M.}~\bibnamefont {Reichert}},\ }\href@noop {} {\  (\bibinfo {year} {2025}{\natexlab{b}})},\ \bibinfo {note} {in preparation}\BibitemShut {NoStop}%
\bibitem [{\citenamefont {Kluth}\ \emph {et~al.}(2023)\citenamefont {Kluth}, \citenamefont {Litim},\ and\ \citenamefont {Reichert}}]{Kluth:2022wgh}%
  \BibitemOpen
  \bibfield  {author} {\bibinfo {author} {\bibfnamefont {Y.}~\bibnamefont {Kluth}}, \bibinfo {author} {\bibfnamefont {D.~F.}\ \bibnamefont {Litim}}, \ and\ \bibinfo {author} {\bibfnamefont {M.}~\bibnamefont {Reichert}},\ }\href {\doibase 10.1103/PhysRevD.107.025011} {\bibfield  {journal} {\bibinfo  {journal} {Phys. Rev. D}\ }\textbf {\bibinfo {volume} {107}},\ \bibinfo {pages} {025011} (\bibinfo {year} {2023})},\ \Eprint {http://arxiv.org/abs/2207.14510} {arXiv:2207.14510 [hep-th]} \BibitemShut {NoStop}%
\bibitem [{\citenamefont {Ihssen}\ and\ \citenamefont {Pawlowski}(2023)}]{Ihssen:2023nqd}%
  \BibitemOpen
  \bibfield  {author} {\bibinfo {author} {\bibfnamefont {F.}~\bibnamefont {Ihssen}}\ and\ \bibinfo {author} {\bibfnamefont {J.~M.}\ \bibnamefont {Pawlowski}},\ }\href@noop {} {\  (\bibinfo {year} {2023})},\ \Eprint {http://arxiv.org/abs/2305.00816} {arXiv:2305.00816 [hep-th]} \BibitemShut {NoStop}%
\bibitem [{\citenamefont {Ihssen}\ and\ \citenamefont {Pawlowski}(2024)}]{Ihssen:2024ihp}%
  \BibitemOpen
  \bibfield  {author} {\bibinfo {author} {\bibfnamefont {F.}~\bibnamefont {Ihssen}}\ and\ \bibinfo {author} {\bibfnamefont {J.~M.}\ \bibnamefont {Pawlowski}},\ }\href@noop {} {\  (\bibinfo {year} {2024})},\ \Eprint {http://arxiv.org/abs/2409.13679} {arXiv:2409.13679 [hep-th]} \BibitemShut {NoStop}%
\bibitem [{\citenamefont {Capper}(1980)}]{Capper:1979ej}%
  \BibitemOpen
  \bibfield  {author} {\bibinfo {author} {\bibfnamefont {D.~M.}\ \bibnamefont {Capper}},\ }\href {\doibase 10.1088/0305-4470/13/1/022} {\bibfield  {journal} {\bibinfo  {journal} {J. Phys. A}\ }\textbf {\bibinfo {volume} {13}},\ \bibinfo {pages} {199} (\bibinfo {year} {1980})}\BibitemShut {NoStop}%
\bibitem [{\citenamefont {fQCD collaboration}()}]{fQCD}%
  \BibitemOpen
  \bibfield  {author} {\bibinfo {author} {\bibnamefont {fQCD collaboration}},\ }\href@noop {} {}\bibinfo {howpublished} {\url{https://fqcd-collaboration.github.io/}}\BibitemShut {NoStop}%
\bibitem [{\citenamefont {Johnson}(2013)}]{quadgk}%
  \BibitemOpen
  \bibfield  {author} {\bibinfo {author} {\bibfnamefont {S.~G.}\ \bibnamefont {Johnson}},\ }\href@noop {} {\enquote {\bibinfo {title} {{QuadGK.jl}: {G}auss--{K}ronrod integration in {J}ulia},}\ }\bibinfo {howpublished} {\url{https://github.com/JuliaMath/QuadGK.jl}} (\bibinfo {year} {2013})\BibitemShut {NoStop}%
\bibitem [{\citenamefont {Johnson}(2017)}]{HCubature}%
  \BibitemOpen
  \bibfield  {author} {\bibinfo {author} {\bibfnamefont {S.~G.}\ \bibnamefont {Johnson}},\ }\href@noop {} {\enquote {\bibinfo {title} {The {HCubature.jl} package for multi-dimensional adaptive integration in {Julia}},}\ }\bibinfo {howpublished} {\url{https://github.com/JuliaMath/HCubature.jl}} (\bibinfo {year} {2017})\BibitemShut {NoStop}%
\bibitem [{\citenamefont {Genz}\ and\ \citenamefont {Malik}(1980)}]{Genz1980}%
  \BibitemOpen
  \bibfield  {author} {\bibinfo {author} {\bibfnamefont {A.~C.}\ \bibnamefont {Genz}}\ and\ \bibinfo {author} {\bibfnamefont {A.~A.}\ \bibnamefont {Malik}},\ }\href {\doibase 10.1016/0771-050x(80)90039-x} {\bibfield  {journal} {\bibinfo  {journal} {Journal of Computational and Applied Mathematics}\ }\textbf {\bibinfo {volume} {6}},\ \bibinfo {pages} {295} (\bibinfo {year} {1980})}\BibitemShut {NoStop}%
\end{thebibliography}%

\end{document}